\def\preprint{1}			% Use for submitted manuscript
\ifdefined\preprint
  \documentclass[preprint,review,12pt,number,sort&compress]{elsarticle}
\fi
\ifdefined\wordcount
  \documentclass[final,3p,times,twocolumn]{elsarticle}
\fi
\ifdefined\final
  \documentclass[final,3p,times,twocolumn]{elsarticle}
\fi

\usepackage{graphicx,stfloats}
\usepackage{color}

\usepackage{blindtext}
\usepackage{latexsym,amsmath,amssymb}
\usepackage[T1]{fontenc}
\usepackage[utf8]{inputenc}
\usepackage[english]{babel} 
\usepackage{textcomp}
\usepackage[normalem]{ulem}
\usepackage{csquotes}

\usepackage{algorithm2e}
\usepackage{scrextend}
\usepackage{mwe,tikz}
\usepackage[percent]{overpic}
\usepackage{color}
\usepackage{siunitx}
\DeclareSIUnit\atm{atm}
\usepackage{graphicx}
\usepackage{tabularx}
\usepackage{subcaption}
\usepackage[export]{adjustbox}
\usepackage{wrapfig}
\graphicspath{{figures/}}
\usepackage[version=4]{mhchem}
\usetikzlibrary{decorations.pathmorphing}
\tikzset{snake it/.style={decorate, decoration=snake}}

\usepackage{booktabs,multirow,multicol}
\usepackage[retainorgcmds]{IEEEtrantools}

\biboptions{sort&compress, square, comma}
\usepackage[breaklinks=true, linkcolor=blue, citecolor=blue, colorlinks=true]{hyperref}

\journal{Combustion and Flame}

\begin{document}

\begin{frontmatter}

\title{Assessing the impact of multicomponent diffusion in direct numerical simulations of premixed, high-Karlovitz, turbulent flames}

\author[1]{Aaron J.~Fillo}

\author[2]{Jason Schlup}
\author[3]{Guillaume Blanquart}
\author[1]{Kyle E.~Niemeyer\corref{cor1}}
\ead{kyle.niemeyer@oregonstate.edu}

\address[1]{School of Mechanical, Industrial, and Manufacturing Engineering, Oregon State University, Corvallis, OR 97331, USA}
\address[2]{Graduate Aerospace Laboratories, California Institute of Technology, Pasadena, USA}
\address[3]{Mechanical Engineering, California Institute of Technology, Pasadena, USA}

\cortext[cor1]{Corresponding author:}

\begin{abstract}
Implementing multicomponent diffusion models in numerical combustion studies is computationally expensive; to reduce cost, numerical simulations commonly use mixture-averaged diffusion treatments or simpler models.
However, the accuracy and appropriateness of mixture-averaged diffusion has not been verified for three-dimensional, turbulent, premixed flames.
In this study we evaluated the role of multicomponent mass diffusion in premixed, three-dimensional high Karlovitz-number hydrogen, \textit{n}-heptane, and toluene flames, representing a range of fuel Lewis numbers.
We also studied a premixed, unstable two-dimensional hydrogen flame due to the importance of diffusion effects in such cases.
Our comparison of diffusion flux vectors revealed differences of \SIrange{10}{20}{\percent} on average between the mixture-averaged and multicomponent diffusion models, and greater than \SI{40}{\percent} in regions of high flame curvature.
Overall, however, the mixture-averaged model produces small differences in diffusion flux compared with global turbulent flame statistics.
To evaluate the impact of these differences between the two models, we compared normalized turbulent flame speeds and conditional means of species mass fraction and source term.
We found differences of \SIrange{5}{20}{\percent} in the mean normalized turbulent flame speeds, which seem to correspond to differences of \SIrange{5}{10}{\percent} in the peak fuel source terms.
Our results motivate further study into whether the mixture-averaged diffusion model is always appropriate for DNS of premixed turbulent flames.
\end{abstract}

\end{frontmatter}

%% Please do not modify the following three lines
\ifdefined \wordcount
\clearpage
\fi

\section{Introduction}\label{Introduction}
Mass, heat, and momentum diffuse simultaneously in turbulent reacting flows, affecting local transport and consumption of chemical species at small time and length scales~\cite{Bird1960,Law2006}.
This coupling of turbulent mixing and heat release during the combustion process can locally impact the flame's structure, curving it and forming steep, multi-directional gradients in the temperature and scalar fields~\cite{peters2001turbulent}.
In these regions of high flame curvature, mass diffusion transport is most accurately represented by the multicomponent diffusion model, which uses a dense matrix of coupled diffusion coefficients to evaluate the relative transport of each chemical species against the remaining species in the mixture~\cite{Hirschfelder1954}.
The Maxwell--Stefan multicomponent diffusion model comes from Boltzmann's kinetic theory of gases~\cite{Curtiss1949TransportMixtures,Furry1948,Hirschfelder1954,Williams1958,VanDeRee1967,Chapman1970,Ferziger1972,CHELLIAH1987,Ern1995,Lam2006MulticomponentRevisited}, and is the most rigorous model for mass diffusion in reacting-flow simulations.

However, modeling full multicomponent mass diffusion transport in a direct numerical simulation (DNS) can be computationally expensive, caused both by the cost of calculating the diffusion coefficients and the memory required to store the multicomponent diffusion coefficient matrix at every location~\cite{Fillo2018}.
As a result, researchers typically use simplified diffusion models to reduce the computational costs associated with calculating the diffusion coefficients~\cite{Xin:2015,Burali2016AssessmentFlows}.
These include, in order of increasing complexity and accuracy, the unity Lewis number, constant non-unity Lewis number~\cite{Poinsot:2005}, and mixture-averaged diffusion assumptions~\cite{Bird1960}.
These models approximate the full multicomponent diffusion coefficient matrix as a constant scalar, a constant vector, and a non-constant diagonal matrix, respectively, reducing the high computational expense associated with numerical combustion studies ~\cite{Burali2016AssessmentFlows,Lapointe2016FuelFlames,Schlup2018ctm,Bird1960,Hirschfelder1954}.
In addition, several approaches further reduce the system's complexity by approximating multicomponent diffusion processes in terms of equivalent Fickian processes, such as those used by Warnatz \cite{Warnatz1978CalculationFlames} and Coltrin et al.~\cite{Coltrin1986ADeposition}. 
While these assumptions may be computationally efficient, to our knowledge, the accuracy and appropriateness of the physics they model has not been evaluated against full multicomponent mass diffusion for DNS of three-dimensional turbulent flames at moderate-to-high Karlovitz numbers (e.g., \numrange{140}{210}).

Although few results exist from three-dimensional reacting flow simulations with multicomponent transport, several studies have investigated the effects of multicomponent transport in simpler configurations.
These studies include one-dimensional~\cite{Coffee:1981,Warnatz:1982, Ern:1999, Bongers:2003, Xin:2015} and two-dimensional flames~\cite{Charentenay:2002,Giovangigli2015MulticomponentFlames} of various unburnt conditions. 
These works compared the multicomponent model with various diffusion and transport property models (from constant Lewis number to mixture-averaged properties).
In general, these studies highlighted the importance of differential diffusion effects but only investigated simplified flame configurations where these effects are relatively small, such as unstretched laminar flames.
% In general, only minor errors exist between multicomponent and mixture-averaged formulations for these one and two dimensional studies, especially in simplified flame configurations, such as unstretched laminar flames.

For example, in evaluating five simplified diffusion models, Coffee and Heimerl~\cite{Coffee:1981} observed that laminar flame speed and species profiles are more sensitive to the input values of individual species transport properties than the specific model used, using simulations of one-dimensional, steady, laminar, premixed hydrogen flames.
They noted that their findings do not indicate that transport phenomenon or model selection are unimportant, but rather that even low-complexity models can be calibrated by carefully selecting the species transport properties to improve accuracy.

Focusing more on the underlying physics of differential diffusion,
Ern and Giovangigli~\cite{Ern:1999} demonstrated that both methane and hydrogen counterflow flames are sensitive to multicomponent transport.
Specifially, neglecting multicomponent effects can lead to overpredicting the extinction strain rate, especially in rich hydrogen flames.
Similarly, Charentenay and Ern~\cite{Charentenay:2002} demonstrated that multicomponent transport only moderately affects global flame properties in two-dimensional, low Karlovitz number, premixed hydrogen/oxygen flames, thanks to the smoothing induced by turbulent fluctuations.
However, when in highly curved flames or flames with local quenching, such as at moderate-to-high Karlovitz numbers, they concluded that the sufficiently large impact of multicomponent transport justifies its inclusion in accurate DNS.

Despite this evidence that multicomponent transport may impact the accuracy of turbulent premixed DNS, studies of three-dimensional turbulent flames continue to rely on simplified diffusion models and do not consider their accuracy relative to multicomponent diffusion, in complex configurations.
Prior evaluations of diffusion models in three-dimensional simulations compared the unity Lewis number, constant but non-unity Lewis number, and mixture-averaged approximations.
For example, Lapointe and Blanquart~\cite{Lapointe2016FuelFlames} compared the relative accuracy of the unity and non-unity Lewis number assumptions for \textit{n}-heptane, iso-octane, toluene, and methane flames.
The flames simulated using the non-unity Lewis number approximation have lower turbulent flame speeds than similar flames simulated with the unity Lewis number assumption.
They attributed these differences to reduced fuel-consumption rates caused by differential diffusion effects~\cite{Lapointe2016FuelFlames}.
Similarly, Burali et al.~\cite{Burali2016AssessmentFlows} compared the non-unity Lewis number assumption to the mixture-averaged diffusion for lean, unstable hydrogen/air flames and lean, turbulent \textit{n}-heptane/air flames.
They demonstrated that using the unity Lewis number assumption underpredicts by \SI{50}{\percent} or more the conditional means of the fuel mass fraction and source term, but using the non-unity Lewis number assumption results in much smaller differences, on the order of \SI{3}{\percent} or less; both were compared with simulations using the mixture-averaged assumption~\cite{Burali2016AssessmentFlows}.
Moreover, Burali et al.~\cite{Burali2016AssessmentFlows} demonstrated that the relative difference associated with the non-unity Lewis number assumption can be minimized by carefully selecting the Lewis-number vector for a wide range of flames, including non-premixed turbulent configurations.

These results reinforce previous conclusions that differential-diffusion effects can impact flame dynamics.
However, there has not been a detailed investigation of the accuracy and appropriateness of the mixture-averaged diffusion model relative to full multicomponent diffusion for turbulent reacting flows.
For high-pressure, non-reacting systems, Borchesi and Bellan~\cite{Borghesi2015iAConditions} developed and analyzed multicomponent species mass flux and turbulent mixing models for large-eddy simulations.
They focused on turbulent mixing of a five-species combustion-relevant mixture of \textit{n}-heptane, oxygen, carbon dioxide, nitrogen, and water.
Their multicomponent transport model significantly improves the accuracy and fidelity of the solution throughout the mixing layer.
However, as this study was restricted to non-reacting flows, it did not assess the impact of multicomponent transport on the chemistry inherent in turbulent combustion.

Motivated by the observed differences between the mixture-averaged and simpler diffusion models, several groups have developed affordable multicomponent transport models.
Ern and Giovangigli~\cite{Ern1995,Ern:1998,Ern:1999} developed the computationally efficient Fortran library EGLIB for accurately determining transport coefficients in gas mixtures.
Ambikasaran and Narayanaswamy~\cite{Ambikasaran2017AnVelocities} proposed an efficient algorithm to compute multicomponent diffusion velocities, which scales linearly with the number of species.
Both methods reduce the computational cost of inverting the dense matrix associated with the Stefan--Maxwell equations~\cite{Bird1960,maxwell1867,stefan1871}.
Most recently, Fillo et al.~\cite{Fillo2018} proposed a fast, semi-implicit, low-memory algorithm for implementing multicomponent mass diffusion, which we use here with the DNS code NGA. 
As a preliminary demonstration of their method, Fillo et al.~\cite{Fillo2018} simulated lean, premixed, three-dimensional turbulent hydrogen/air flames at moderate-to-high Karlovitz numbers using the mixture-averaged and multicomponent diffusion models.
In these flames, the mixture-averaged diffusion model underpredicts the peak mean source term and normalized turbulent flame speed by \SI{5.5}{\percent} and \SI{15}{\percent}, respectively~\cite{Fillo2018}.

In addition to mass diffusion, several groups have also investigated the impact of multicomponent Soret and Dufour thermal diffusion effects.
In particular, studies have examined the importance of including thermal diffusion in a wide range of flame configurations \cite{Coffee:1981,Ern:1998,Ern:1999,Bongers:2003,Yang2010,Xin2012,Giovangigli2015MulticomponentFlames,Schlup2018ctm,Han2020}.
For example, Giovangigli \cite{Giovangigli2015MulticomponentFlames} demonstrated that multicomponent Soret effects significantly impact a wide range of laminar hydrogen/air flames: laminar flame speeds in flat flames and extinction stretch rates in strained premixed flames.
Using a mechanistic approach, Yang et al.~\cite{Yang2010} observed that Soret diffusion of hydrogen radical (\ce{H}) in premixed hydrogen flames actively modifies its concentration and distribution in the reaction zone.
This effect was especially evident in symmetric, twin counter-flow, premixed hydrogen flames, where Soret diffusion increases and decreases individual reaction rates in lean and rich mixtures, respectively.
Performing a similar mechanistic approach examining planar and stretched premixed \textit{n}-heptane and hydrogen flames, Xin et al.~\cite{Xin2012} demonstrated that these chemical kinetic effects result from Soret diffusion diluting or enriching the reactant concentrations in the reaction front, and could substantially impact fuel burning rates---especially in highly stretched flames.
Han et al.~\cite{Han2020} recently examined Soret diffusion in turbulent non-premixed hydrogen flames, comparing DNS using mixture-averaged diffusion with and without Soret effects.
They found that it significantly affects \ce{H} and \ce{OH} profiles in the flame but negligibly modifies \ce{H2}.

Finally, Schlup and Blanquart~\cite{Schlup2018ctm} examined the impact of multicomponent thermal diffusion in DNS of turbulent, premixed, high-Karlovitz hydrogen/air flames.
They observed that simulations using the mixture-averaged thermal diffusion assumption underpredict flame speeds compared with simulations using full multicomponent thermal diffusion.
In addition, they observed that including multicomponent thermal diffusion increases local production rates in in regions of high positive curvature~\cite{Schlup2018ctm}.
These observed discrepancies in similar flame simulations with different diffusion models warrant a detailed investigation of the fundamental transport phenomena involved.
However, while thermal diffusion can be important in some fuel/air mixtures, in this article we focus on mass diffusion, and direct interested readers to the work of Schlup and Blanquart~\cite{Schlup2018ctm}, for example, for an investigation of these effects.

The primary objective of this study is to evaluate the accuracy and appropriateness of the mixture-averaged diffusion assumption for use in DNS of premixed unsteady laminar and turbulent flames.
This objective will be realized via an a priori analysis of the orientation and magnitude of the mixture-averaged diffusion flux vector, relative to that of the multicomponent model, for a range of flame configurations.
We will further analyze differences between the diffusion models by considering a posteriori results of turbulent flame structures (i.e., species mass fraction and source term profiles).
Finally, we will compare the time history and average normalized turbulent flame speeds of hydrogen/air, \textit{n}-heptane/air, and toluene/air flames as a global measure of the differences between the multicomponent and mixture-averaged diffusion models.

The paper is organized as follows: 
Section~\ref{sec:numericalapproach} describes the governing equations, diffusion models, and flow configurations for the simulations.
Then, Section~\ref{sec:results} presents and discusses the results from a priori, a posteriori, turbulent flame speed, and chemical pathway analyses.
Finally, in Section~\ref{sec:conclusions} we draw conclusions from the comparisons of the diffusion models.

\section{Numerical approach}\label{sec:numericalapproach}
This section describes the governing reacting-flow equations and flow solver used, and briefly discusses the diffusion models to be studied. 
It also presents the two- and three-dimensional flow configurations used.

\subsection{Governing equations}

We solve the variable-density, low Mach number, reacting flow equations using
the finite-difference code NGA \cite{Desjardins2008,Savard2015AChemistry}.
The complete conservation equations are
{\allowdisplaybreaks \begin{IEEEeqnarray}{rCl}
\frac{\partial\rho}{\partial t}+\nabla\cdotp(\rho\textbf{u}) &=& 0 \;, 
\label{eq:continuity} \\
\frac{\partial\rho \textbf{u}}{\partial t}+\nabla\cdotp(\rho\textbf{ u}\otimes\textbf{u}) &=& -\nabla p+\nabla\cdotp\boldsymbol{\tau}+\textbf{f} \;,
\label{eq:momentum} \\
\frac{\partial\rho T}{\partial t}+\nabla\cdotp(\rho \textbf{u}T) &=& \nabla\cdotp(\rho\alpha\nabla T)+\rho\dot{\omega}_{T} - \frac{1}{c_{p}}\sum_{i}^N c_{p,i}\textbf{j}_{i} \cdotp \nabla T+\frac{\rho\alpha}{c_{p}}\nabla{c_{p}}\cdotp\nabla T \;,
\label{eq:energy} \IEEEeqnarraynumspace \\
\frac{\partial\rho Y_{i}}{\partial t} + \nabla\cdotp(\rho \textbf{u} Y_{i}) &=& -\nabla\cdotp \textbf{j}_{i}+ \dot{\omega_{i}} \;,
\label{eq:species}
\end{IEEEeqnarray}}%
where $\rho$ is the mixture density, $t$ is time, $\textbf{u}$ is the velocity, $p$ is the hydrodynamic pressure, $\boldsymbol{\tau}$ is the viscous stress tensor, $\textbf{f}$ represents volumetric forces, $T$ is the temperature, $\alpha$ is the mixture thermal diffusivity, $c_{p,i}$ is the constant-pressure specific heat of species $i$, $N$ is the number of species, $c_{p}$ is the constant-pressure specific heat of the mixture, and $\textbf{j}_{i}$, $Y_{i}$ , and $\dot{\omega_{i}}$ are the diffusion flux, mass fraction, and production rate of species $i$, respectively.
In Eq.~\eqref{eq:energy}, the temperature source term is given by
\begin{equation} \label{eq:tempsource}
\dot{\omega}_{T} = \frac{-1}{c_p} \sum_{i}^N h_{i}(T)\dot{\omega_{i}} \;,
\end{equation}
where $h_{i}(T)$ is the specific enthalpy of species $i$ as a function of temperature.
The density is determined from the ideal gas equation of state.

NGA solves Eqs.\eqref{eq:continuity}--\eqref{eq:species} using a numerical scheme second-order accurate in both space and time~\cite{Desjardins2008,Savard2015AChemistry}, via a semi-implicit Crank--Nicolson time integration method~\cite{Pierce2001}. 
It uses the third-order Bounded QUICK scheme (BQUICK)~\cite{Herrmann2006} for scalar transport.
We discuss the diffusion solver in more detail next in Section~\ref{sec:diffusion_models}.

\subsection{Overview of diffusion models}\label{sec:diffusion_models}
The diffusion fluxes are calculated using the semi-implicit scheme developed by Fillo et al.~\cite{Fillo2018} with either mixture-averaged or multicomponent \cite{Hirschfelder1954,Bird1960} models, both of which are based on Boltzmann's equation for the kinetic theory of gases \cite{Curtiss1949TransportMixtures, Hirschfelder1954}.
For this study, we neglect both baro-diffusion and thermal diffusion (Soret and Dufour effects).
The baro-diffusion term is commonly neglected in reacting-flow simulations under the low Mach number approximation \cite{Grcar2009TheFlames}.
We also neglect thermal diffusion because our objective is to investigate the impact of mass diffusion models; Schlup and Blanquart previously explored the effects of thermal diffusion modeling~\cite{Schlup2018ctm}.

The species diffusion flux for the mixture-averaged diffusion model (abbreviated by MA hereafter) is related to the species gradient by a Fickian formulation, and is expressed as
\begin{equation} \label{eq:MA-diffusion-flux}
\textbf{j}_{i}^{\text{MA}} = -\rho D_{i, m}\frac{Y_i}{X_{i}}\nabla X_{i}+\rho Y_{i}\textbf{u}'_{c} \;,
\end{equation}
where $X_i$ is the $i$th species mole fraction, $D_{i,m}$ is the $i$th species mixture-averaged diffusion coefficient as expressed by Bird et al.~\cite{Bird1960}:
\begin{equation} \label{eq:MA-diffusion-coefficient}
D_{i, m} = \frac{1-Y_{i}}{\sum_{j\neq i}^{N} X_{j}/\mathcal{D}_{ji}} \;,
\end{equation}
where $\mathcal{D}_{ji}$ is the binary diffusion coefficient of species $i$ and $j$, and $\textbf{u}'_{c}$ is the correction velocity used to ensure mass continuity:
\begin{equation} \label{eq:correction-velocity}
\textbf{u}'_{c} = \sum_{i}^N D_{i, m}\frac{Y_{i}}{X_{i}}\nabla X_{i} \;.
\end{equation}

Alternatively, the multicomponent diffusion model (abbreviated as MC hereafter), as presented by Bird et al.~\cite{Bird1960}, calculates the species diffusion flux as
\begin{equation} \label{eq:MC-diffusion-flux}
\textbf{j}_{i}^{\text{MC}}=\frac{\rho Y_{i}}{X_{i}W}\sum_{j}^{N} W_{j}D_{ij}\nabla{X_{j}} \;,
\end{equation}
where $W$ is the mixture molecular weight, $W_{j}$ is the molecular weight of the $j$th species, and $D_{ij}$ is the ordinary multicomponent diffusion coefficient between species $i$ and $j$, which we compute here using the \texttt{MCMDIF} subroutine of CHEMKIN II~\cite{Kee1989Chemkin-II:Kinetics} with the method outlined by Dixon--Lewis~\cite{Dixon-Lewis1968FlameSystems}.

Though Fillo et al.~\cite{Fillo2018} provide further details on how these methods are implemented in NGA, we will summarize the key aspects here.
First, we modified the treatment of mass-diffusion terms in the semi-implicit scheme of Savard et al.~\cite{Savard2015AChemistry}, using the mixture-averaged diffusion coeﬃcient matrix to precondition the diffusion source term.
Furthermore, the multicomponent implementation uses a dynamic memory algorithm to reduce the number of times the multicomponent diffusion coeﬃcient matrix must be evaluated.
Fillo et al.\ showed the stability and accuracy of this semi-implicit scheme~\cite{Fillo2018},
and that the computational costs of the mixture-averaged and multicomponent diffusion models scale linearly and quadratically with the number of species in the chemical kinetic model.
Due to the efficient memory algorithm, most of the cost comes from the Chemkin II~\cite{Kee1989Chemkin-II:Kinetics} routines used to calculate the diffusion coefficients.

\begin{table}[htbp]
\scriptsize
    \caption{Parameters of the two- and three-dimensional simulations.  
    $\Delta x$ is the grid spacing, $\eta_{u}$ is Kolmogorov length scale in the unburnt gas, $\Delta t$ is the simulation time step, $\phi$ is the equivalence ratio, $p_0$ is the thermodynamic pressure, $T_u$ is the temperature of the unburnt mixture, $T_{\text{peak}}$ is the temperature of peak fuel consumption rate in the one-dimensional laminar flame, $S_L$ is the laminar flame speed, $l_F = \left(T_b - T_u\right)/\left|\nabla T\right|_{\text{max}}$ is the laminar flame thickness, $l = u'^3/\epsilon$ is the integral length scale, $u'$ is the turbulence fluctuations, $\epsilon$ is the turbulent energy dissipation rate, $\text{Ka}_u$ is the Karlovitz number in the unburnt mixture, $\text{Re}_t$ is the turbulent Reynolds number in the unburnt mixture, $\nu_u$ is the unburnt kinematic viscosity, $A_{\text{force}}$ is the turbulent forcing coefficient used in NGA~\cite{Carroll2013}, and $\text{Le}_F$ is the fuel Lewis number, where $D_F$ is the fuel diffusion coefficient from the mixture-averaged model.}
    \centering
    %\begin{tabular}{@{\extracolsep{\fill}}l c c c c c c c c@{}}
    \begin{tabular}{@{}l c c c c c c c c@{}}
         \toprule
         & \multicolumn{2}{c}{\ce{H2} (2D)} & \multicolumn{2}{c}{\ce{H2}} & \multicolumn{2}{c}{\textit{n}-\ce{C7H16}} & \multicolumn{2}{c}{\ce{C6H5CH3}} \\
         \midrule
         & MA & MC & MA & MC & MA & MC & MA & MC \\
         \midrule
         Domain & \multicolumn{2}{c}{$4L \times L$}  & \multicolumn{2}{c}{$8L \times L \times L$} & \multicolumn{2}{c}{$11L \times L \times L$} & \multicolumn{2}{c}{$11L \times L \times L$} \\
         $L$ & \multicolumn{2}{c}{$472\Delta{x}$} & \multicolumn{2}{c}{$190\Delta{x}$} & \multicolumn{2}{c}{$128\Delta{x}$} & \multicolumn{2}{c}{$128\Delta{x}$} \\
         Grid & \multicolumn{2}{c}{$1888\times472$}  & \multicolumn{2}{c}{$1520\times190\times190$} & \multicolumn{2}{c}{$1408\times128\times128$} & \multicolumn{2}{c}{$1408\times128\times128$} \\
         $\Delta{x}$ [\si{\meter}] & \multicolumn{2}{c}{\num{4.24e-5}} & \multicolumn{2}{c}{\num{4.24e-5}} & \multicolumn{2}{c}{\num{1.8e-5}} & \multicolumn{2}{c}{\num{1.8e-5}} \\
         $\eta_{u}$ [\si{\m}] & \multicolumn{2}{c}{--} & \multicolumn{2}{c}{\num{2.1e-5}} & \multicolumn{2}{c}{\num{9.0e-6}} & \multicolumn{2}{c}{\num{9.1e-6}} \\
         $\Delta{t}$ [\si{\s}] & \multicolumn{2}{c}{\num{5e-6}} & \multicolumn{2}{c}{\num{6e-7}} & \multicolumn{2}{c}{\num{6e-7}} & \multicolumn{2}{c}{\num{6e-7}} \\
         $\phi$ & \multicolumn{2}{c}{0.4} & \multicolumn{2}{c}{0.4} & \multicolumn{2}{c}{0.9} & \multicolumn{2}{c}{0.9} \\
         $p_0$ [\si{\atm}] & \multicolumn{2}{c}{1} & \multicolumn{2}{c}{1} & \multicolumn{2}{c}{1} & \multicolumn{2}{c}{1}\\
         $T_{\text{u}}$ [\si{\K}] & \multicolumn{2}{c}{298} & \multicolumn{2}{c}{298} & \multicolumn{2}{c}{298} & \multicolumn{2}{c}{298}\\
         $T_{\text{peak}}$ [\si{\K}] & 1190 & 1180 & 1190 & 1180 & 1270 & 1230 & 1420 & 1420 \\
         $S_{L}$ [\si{\centi\meter\per\second}] & 23.0 & 22.3 & 23.0 & 22.3 & 35.1 & 37.3 & 34.3 & 34.4 \\
         $l_{F}$ [\si{\mm}] & 0.643 & 0.651 & 0.643 & 0.631 & 0.390 & 0.385 & 0.410 & 0.420 \\
         $l/l_{F}$ & \multicolumn{2}{c}{--} & 2 & 2.04 & \multicolumn{2}{c}{1.1} & \multicolumn{2}{c}{1.1} \\
         $u'/S_{L}$ & \multicolumn{2}{c}{--} & 18 & 18.6 & 18 & 16.9 & 17 & 16.9 \\
         $\text{Ka}_u = \tau_{F}/\tau_{\eta}$ & \multicolumn{2}{c}{--} & 149 & 151 & 220 & 207 & 200 & 204 \\
         $\text{Re}_t = (u'l)/\nu_{u}$ & \multicolumn{2}{c}{--} & \multicolumn{2}{c}{289} & \multicolumn{2}{c}{190} & \multicolumn{2}{c}{175}  \\
         $A_{\text{force}}$ [\si{\per\second}]& \multicolumn{2}{c}{--} & \multicolumn{2}{c}{973.05} & \multicolumn{2}{c}{4730} & \multicolumn{2}{c}{4333}  \\
         $\text{Le}_F = \alpha\slash D_F$ & \multicolumn{2}{c}{0.3} & \multicolumn{2}{c}{0.3} & \multicolumn{2}{c}{2.8} & \multicolumn{2}{c}{2.5} \\
         \bottomrule
    \end{tabular}
    \label{tab:2.3D_flow_config}
\end{table}

\subsection{Flow configuration}

We used three flow configurations in this work.
The first is a one-dimensional, unstretched (flat), laminar, hydrogen/air flame with an unburnt temperature of \SI{298}{\kelvin}, pressure of \SI{1}{atm}, and equivalence ratio of $\phi=0.4$.
To ensure the flame remained centered in the computational domain, comprised of 720 grid points where $\Delta x=$\SI{15.4}{\micro\meter}, we set the inlet velocity as equal to the laminar flame speed:
\begin{equation}
S_{L} = -\frac{\int \rho\dot{\omega}_{\ce{H2}}dx}{\rho_{u}Y_{\ce{H2},u}} \;,
\end{equation}
where $\rho_{u}$ is the unburnt mixture density and $Y_{\ce{H2},u}$ is the unburnt fuel mass fraction.
We selected the grid spacing to ensure at least 20 points through the laminar flame, with the thickness defined using
the maximum temperature gradient: $l_{F} = (T_{\max} - T_{\min})/\lvert\nabla{T\rvert_{\max}}$.
Schlup and Blanquart~\cite{Schlup2018ctm} used an identical configuration to investigate the impact of Soret and Dufour thermal diffusion effects.

For the second and third configurations, we selected multidimensional cases where diffusion modeling may be particularly important.
The second configuration considered is a two-dimensional domain used to study unsteady, freely propagating lean hydrogen/air flames \cite{Burali2016AssessmentFlows,Schlup2018ctm}.
The third configuration is a doubly-periodic, turbulence-in-a-box configuration we used to study three-dimensional statistically stationary hydrogen/air, n-heptane/air, and toluene/air flames \cite{Burali2016AssessmentFlows,Lapointe2016FuelFlames,Schlup2018ctm}.
The selected fuels also span a range of Lewis numbers: $\text{Le}_{\ce{H2}} = 0.3$, $\text{Le}_{\ce{C6H5CH3}} = 2.5$, and $\text{Le}_{\ce{C7H16}} = 2.8$.
This allows us to evaluate if the relative strength of mass diffusivity relative to thermal diffusivity affects a flame's sensitivity to multicomponent diffusion.
For example, the low Lewis number of hydrogen can result in differential diffusion effects, which cause the instabilities found in lean hydrogen/air flames.
Further, we selected the turbulence timescales at the high Karlovitz numbers considered to match the order of magnitude of the diffusion time scales, such that diffusion may interact with turbulence.
All three configurations have been used in previous studies, and so we provide only a brief overview here.

\subsubsection{Two-dimensional flow configuration}
\label{sec:twoDconfig}

% A similar configuration has been used to investigate two-dimensional freely-propagating flames~\cite{Grcar2009TheFlamesb,Bastiaans2012NumericalCombustion,Regele2013AFlames}, while this exact configuration was used in~\cite{Burali2016AssessmentFlows,Schlup2018ctm}.  
% Only a brief discussion is given here.
%The flame is initialized with an array of solutions corresponding to the one-dimensional flat flame simulations run previously, with an equivalence ratio of $\phi=0.4$ and an inlet temperature and pressure of \SI{298}{\kelvin} and \SI{1}{atm}, respectively.
% The flame profile is initially perturbed in the span-wise direction by two sinusoidal modes.
The two-dimensional analysis is performed using a hydrogen/air mixture with a nine-species, 54-reaction chemistry model from Hong et al.~\cite{Hong2011AnMeasurements,Lam2013AAbsorption,Hong2013OnAbsorption} (forward and backward reactions are counted separately).
The domain has inlet and convective outlet boundaries in the streamwise direction and periodic boundaries in the spanwise direction.
The inlet velocity boundary condition is fixed at the mean effective burning velocity, such that the unstable flame remains statistically stationary in the domain.  The mean effective burning velocity, $S_{\text{eff}}^{2D}$, is defined as
\begin{equation}\label{eq:flamespeed}
S_{\text{eff}}^{2D}=-\frac{\int_{A} \rho \dot{\omega}_{\ce{H2}} dA}{\rho_{u} Y_{\ce{H2},u} L} \;,
\end{equation}
where $L$ is the spanwise dimension of the computational domain.
This velocity boundary condition allows the simulation to run for an arbitrary length of time to collect statistics.

\begin{figure}[htbp]
\centering
\includegraphics[width=0.95\textwidth]{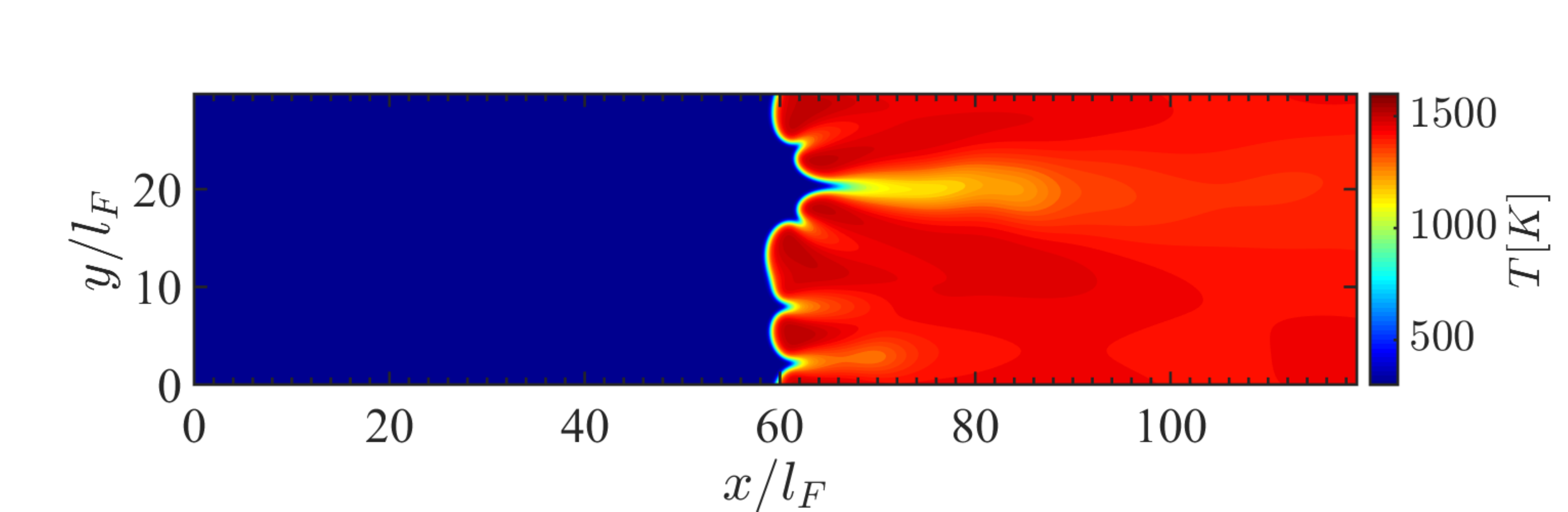}
\caption{Temperature contour for the two-dimensional freely propagating unsteady hydrogen/air flame obtained with the multicomponent diffusion model.} 
\label{2D_temp}
\end{figure}

Table~\ref{tab:2.3D_flow_config} includes details of the computational domain.
%The physical size of the domain is approximately \SI{120}{\textit{l}_{\textit{F}}} in the stream-wise direction by \SI{30}{\textit{l}_{\textit{F}}} in the spanwise direction.
%The grid is a structured, uniform mesh with $1888 \times 472$ cells, with a cell size corresponding to $\Delta x = \Delta y = l_{F}/16$.
The unburnt mixture has an equivalence ratio of $\phi = 0.4$, temperature of $T_u = $ \SI{298}{\kelvin}, and pressure of $p_o = $ \SI{1}{\atm}.
The initial scalar and velocity fields are generated by perturbing a flat, two-dimensional flame profile, using two sinusoidal modes defined by
\begin{equation}
    x_{F,0} = E + A \sum_{i=1,2}\text{cos}\left(2\pi k_i \frac{y}{H}\right)
\end{equation}
where $x_{F,0}$ is the initial flame position, $E$ is the average flame position, $A = \SI{e-4}{\meter}$ is the amplitude, $k_1 = 20$ and $k_2 = 30$ are two
coprime modes, $y$ is the vertical coordinate, and $H$ is the height of the domain~\cite{Burali2016AssessmentFlows}.
Schlup et al.~\cite{Schlup2018ctm} and Burali et al.~\cite{Burali2016AssessmentFlows} used the same set of disturbance parameters to to initially perturb the flame asymmetrically and trigger Darrieus--Landau instabilities.
Figure~\ref{2D_temp} shows an example temperature contour with a representative unsteady flame clearly visible.

\subsubsection{Three-dimensional flow configuration}\label{sec:threeDconfig}

Three fuel/air mixtures are simulated in the three-dimensional configuration: $\phi=0.4$ hydrogen/air, $\phi=0.9$ \textit{n}-heptane/air, and $\phi=0.9$ toluene/air.
The hydrogen/air mixture uses the same chemical-kinetic model as in the two-dimensional case~\cite{Hong2011AnMeasurements,Lam2013AAbsorption,Hong2013OnAbsorption}.
The \textit{n}-heptane/air mixture uses the reduced kinetic model described by Savard et al.~\cite{Savard2015structure,Savard2015} consisting of 35 species and 217 reactions.
Finally, the toluene/air mixture uses the 47-species, 290-reaction kinetic model of Bisetti et al.~\cite{Bisetti2012}.

Table~\ref{tab:2.3D_flow_config} gives the details of the computational domains used for the three-dimensional simulations.
The domains consist of inflow and convective outflow boundary conditions in the streamwise direction, and periodic boundaries in the two spanwise directions.
The inflow velocity is the mean turbulent flame speed, which keeps the flame statistically stationary such that turbulent statistics can be collected over an arbitrarily long run time.
In the absence of mean shear, a linear turbulence forcing method \cite{Rosales2005,Carroll2013} maintains the production of turbulent kinetic energy through the flame.

Table~\ref{tab:2.3D_flow_config} also provides details on the unburnt mixture, corresponding one-dimensional flames, and inlet turbulence.
The unburnt temperatures and pressures for all cases are \SI{298}{\kelvin} and \SI{1}{\atm}, respectively.
Table~\ref{tab:2.3D_flow_config} gives the definitions of the Karlovitz number, $\text{Ka}_u$, and turbulent Reynolds number, $\text{Re}_t$, where $\tau_F = l_F / S_L$ is the flame time scale and $\tau_{\eta}=\left(\nu_u / \epsilon\right)^{1/2}$ is the Kolmogorov time scale of the incoming turbulence.

\section{Results and discussion}\label{sec:results}

This section presents a priori results for the one-dimensional flat hydrogen/air flame and two-dimensional, unsteady, premixed hydrogen/air flame, as well as a priori and a posteriori results for the three-dimensional, turbulent, premixed fuel/air flames of hydrogen, \textit{n}-heptane, and toluene.

\subsection{A priori diffusion flux comparison}\label{sec:aprioriresults}
Mass, momentum, and heat diffusion are strongly coupled processes in reacting flows, and as a result isolating the causes of observed effects can be difficult.
To overcome this challenge we compared the diffusion models with an ``a priori'' analysis that calculates the mass-diffusion flux vectors for each method using identical scalar gradient fields.
This analysis highlights the differences in diffusion flux vectors from each method before any differences influence the flowfield.
By calculating the mass diffusion fluxes in this way, we isolate the effects of the diffusion model on the resulting diffusion vectors from any time evolution of the reacting flow field.
To assess disagreement between the mixture-averaged and multicomponent diffusion models, we evaluate the relative orientation and magnitude of the diffusion flux vectors they produce.

\subsubsection{One-dimensional flame}

\begin{figure}[htbp] 
  \centering
    % \begin{subfigure}{0.3\textwidth}
    %     \includegraphics[width=\textwidth]{1D_flux_N2.pdf}
    %     \caption{\ce{N2}}
    %     \label{1D_flux (a)}
    % \end{subfigure}
    % \begin{subfigure}{0.3\textwidth}
    %     \includegraphics[width=\textwidth]{1D_flux_H.pdf}
    %     \caption{\ce{H}}
    % \end{subfigure}
    % \begin{subfigure}{0.3\textwidth}
    %     \includegraphics[width=\textwidth]{1D_flux_O2.pdf}
    %     \caption{\ce{O2}}
    % \end{subfigure}
    
    % \begin{subfigure}{0.3\textwidth}
    %     \includegraphics[width=\textwidth]{1D_flux_O.pdf}
    %     \caption{\ce{O}}
    % \end{subfigure}
    % \begin{subfigure}{0.3\textwidth}
    %     \includegraphics[width=\textwidth]{1D_flux_OH.pdf}
    %     \caption{\ce{OH}}
    % \end{subfigure}
    % \begin{subfigure}{0.3\textwidth}
    %     \includegraphics[width=\textwidth]{1D_flux_H2.pdf}
    %     \caption{\ce{H2}}
    % \end{subfigure}
    
    % \begin{subfigure}{0.3\textwidth}
    %     \includegraphics[width=\textwidth]{1D_flux_H2O.pdf}
    %     \caption{\ce{H2O}}
    % \end{subfigure}
    % \begin{subfigure}{0.3\textwidth}
    %     \includegraphics[width=\textwidth]{1D_flux_HO2.pdf}
    %     \caption{\ce{HO2}}
    % \end{subfigure}
    % \begin{subfigure}{0.3\textwidth}
    %     \includegraphics[width=\textwidth]{1D_flux_H2O2.pdf}
    %     \caption{\ce{H2O2}}
    % \end{subfigure}
  \includegraphics[width=\textwidth]{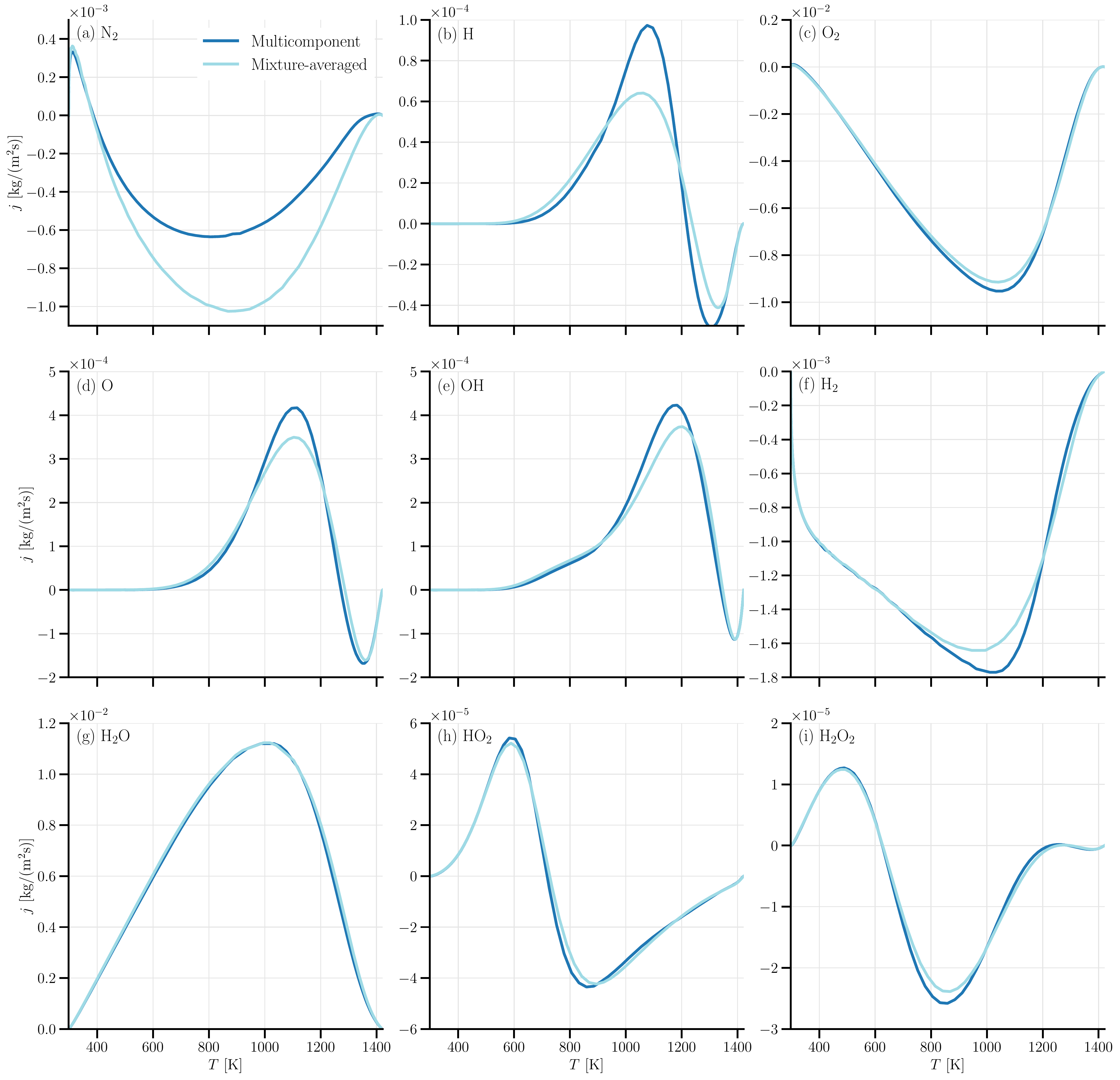}
  \caption{A priori comparisons of mass diffusion fluxes vs.\ temperature in a one-dimensional hydrogen/air flame at $\phi=0.4$. }
  \label{1D_flux}
\end{figure}

Figure~\ref{1D_flux} compares the a priori diffusion fluxes for the one-dimensional, flat, hydrogen/air flame relative to the local mixture temperature.
As expected, the flux profiles for the mixture-averaged and multicomponent cases match in shape and magnitude.
However, the mixture-averaged model underpredicts the maximum flux magnitude of hydrogen radical (\ce{H}) by approximately \SI{40}{\percent}.
Similarly, the mixture-averaged model underpredicts molecular hydrogen (\ce{H2}) and hydroxyl radical (\ce{OH}) fluxes by approximately \SI{18}{\percent}, and oxygen radical (\ce{O}) by \SI{16}{\percent}.
% However, the multicomponent case has an approximately \SI{40}{\percent} higher maximum magnitude for hydrogen radical (\ce{H}) and an approximately \SI{18}{\percent} higher maximum magnitude for molecular hydrogen (\ce{H2}) and hydroxyl radical \ce{OH} fluxes compared with the mixture-averaged case.
These differences are substantial but agree with previous results for one-dimensional premixed hydrogen/air flames~\cite{Ern:1999,Charentenay:2002}. 

These differences disrupt mass continuity by locally altering the equivalence ratio in regions of high mass-diffusion flux.
This effect is clear when considering the correction velocity, which is based on the mole and mass fractions of the species.
As a result, it lumps a large portion of the correction for mass continuity into the \ce{N2} mass flux, which the mixture-averaged model overpredicts by \SI{40}{\percent}.
The correction velocity is not correcting for the errors in the mixture-averaged model; rather, it simply ensures mass continuity.
% This weighting of the correction velocity corrects for the observed differences in the flux magnitude for the other species in the mixture-averaged model to maintain mass continuity, however, errors.
% The maximum magnitude of the \ce{N2} mass flux is approximately \SI{40}{\percent} higher for the mixture-averaged model; this overprediction of nitrogen mass flux corresponds to the observed underprediction of the mixture-averaged model for the remaining species fluxes and fulfills mass-continuity requirements.
% As a secondary check, the species mass fluxes for both models sum to zero, confirming that mass is conserved.

\subsubsection{Multi-dimensional flames}

We next performed an a priori assessment of the species mass diffusion fluxes in the multi-dimensional flames.
% However, based on the mathematical definitions of the two diffusion models, 
However, because of the added degrees-of-freedom in multi-dimensional flows, we now investigate the relative angles of the flux vectors with respect to the species gradient vectors to assess the relative direction of mass flux, in addition to the flux magnitudes. 
The mixture-averaged flux vector for a given species is based on the gradient of that species and, as a result, should align almost perfectly with its gradient and in the opposite direction.
However, some misalignment may also arise because of the velocity correction term in Eq.~\eqref{eq:MA-diffusion-coefficient}.
In contrast, as shown by Eq.~\eqref{eq:MC-diffusion-flux}, the multicomponent flux of a given species is based on the net influence of the remaining species (but not itself) and thus may not necessarily align with its own gradient.
Differential diffusion may misalign the species gradient and multicomponent diffusion flux vectors in regions of high flame curvature where strong gradients can exist in multiple directions.

\begin{figure}[htbp]
    \begin{subfigure}{0.48\textwidth}
        \centering
            \includegraphics[width=\textwidth]{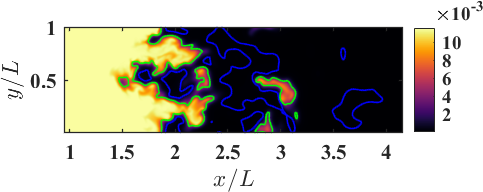}
        \caption{MA $\text{Y}_{\ce{H_2}}$}\label{H2_contours (a)}
    \end{subfigure}
    \begin{subfigure}{0.48\textwidth}
        \centering
            \includegraphics[width=\textwidth]{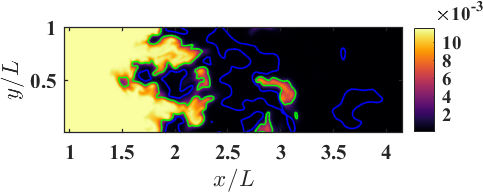}
        \caption{MC $\text{Y}_{\ce{H_2}}$}\label{H2_contours (b)}
    \end{subfigure}\\
    % \begin{subfigure}{0.45\textwidth}
    %     \centering
    %         \includegraphics[width=\textwidth]{H2_log_Y_MA_fulldomain.png}
    %     \caption{MA $Y_{H_2}$}
    % \end{subfigure}
    % \begin{subfigure}{0.45\textwidth}
    %     \centering
    %         \includegraphics[width=\textwidth]{H2_log_Y_MC_fulldomain.png}
    %     \caption{MC $Y_{H_2}$}
    % \end{subfigure}\\
    \begin{subfigure}{0.48\textwidth}
        \centering
            \includegraphics[width=\textwidth]{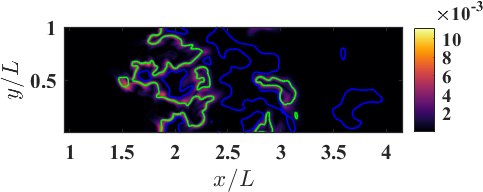}
        \caption{MA $\text{j}_{\ce{H2}}$ [\si{\kilo\gram\per\square\meter\per\second}]}\label{H2_contours (c)}
    \end{subfigure}
    \begin{subfigure}{0.48\textwidth}
        \centering
            \includegraphics[width=\textwidth]{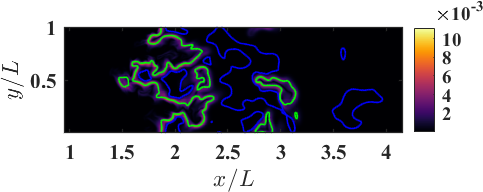}
        \caption{MC $\text{j}_{\ce{H2}}$ [\si{\kilo\gram\per\square\meter\per\second}]}\label{H2_contours (d)}
    \end{subfigure}\\
    \begin{subfigure}{0.48\textwidth}
        \centering
            \includegraphics[width=\textwidth]{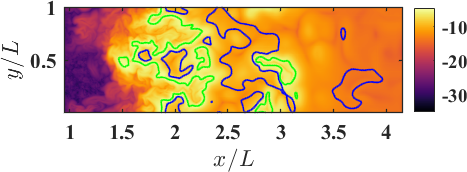}
        \caption{MA $\log_{10}(\text{j}_{\ce{H2}})$}\label{H2_contours (e)}
    \end{subfigure}
    \begin{subfigure}{0.48\textwidth}
        \centering
            \includegraphics[width=\textwidth]{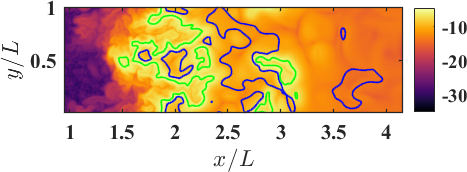}
        \caption{MC $\log_{10}(\text{j}_{\ce{H2}})$}\label{H2_contours (f)}
    \end{subfigure}\\
    \begin{subfigure}{0.48\textwidth}
        \centering
            \includegraphics[width=\textwidth]{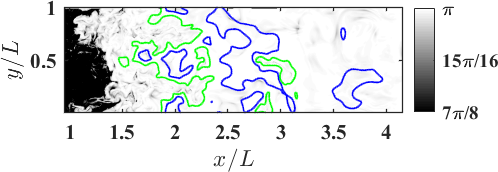}
        \caption{MA $\nabla\text{Y}_{\ce{H2}}\angle\text{j}_{\ce{H_2}}$}\label{H2_contours (g)}
    \end{subfigure}
    \begin{subfigure}{0.48\textwidth}
        \centering
            \includegraphics[width=\textwidth]{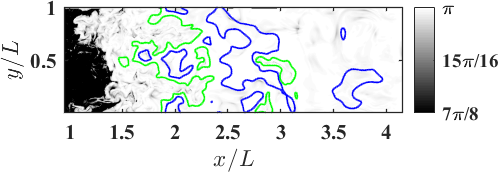}
        \caption{MC $\nabla\text{Y}_{\ce{H2}}\angle\text{j}_{\ce{H_2}}$}\label{H2_contours (h)}
    \end{subfigure}
  \caption{Fields of fuel mass fraction (a, b), fuel mass diffusion flux (c, d), log of the fuel mass diffusion flux (e, f), and angle between fuel mass flux and species gradient vectors (g, h) for one time step of the hydrogen-air turbulent premixed flame for the mixture-averaged (MA) and multicomponent (MC) diffusion cases. 
  Shown are domain cross-sections through the midplane. 
  The green and blue lines correspond to isosurfaces of $T_u=T_{\text{peak}}-\SI{300}{\kelvin}$ and $T_b=T_{\text{peak}}+\SI{300}{\kelvin}$, respectively, and represent the inflow and outflow surfaces of the flame front.}
  \label{H2_contours}
\end{figure}

As a qualitative assessment, Figure~\ref{H2_contours} shows two-dimensional slices of fuel mass fraction, fuel mass diffusion flux, and angle between species flux vector and gradient vector from the turbulent hydrogen/air flame for the mixture-averaged and multicomponent models.
For this assessment an angle of $\pi$ means that the species flux and gradient vectors align, while smaller angles show misalignment.
To help highlight small differences between the two diffusion models, we also present the logarithm of the mass diffusion flux field.
The location of the flame is indicated by isolines of $T = T_{\text{peak}}- $\SI{300}{\kelvin} (green) and $T = T_{\text{peak}}+$ \SI{300}{\kelvin} (blue) included on the fields of fuel mass fraction.

Qualitatively, the angles shown in Figures~\ref{H2_contours (g)} and~\ref{H2_contours (h)} show good agreement between the mixture-averaged and multicomponent models.
At the inlet of the domain (the left side), the relative angle is zero.
In this region fuel has not been consumed, so the fuel mass fraction is approximately constant and gradients are small; as a result the magnitude of the mass flux in this region is nearly zero.
This is confirmed in Figures~\ref{H2_contours (e)} and \ref{H2_contours (f)} where values of $\log_{10}(\text{j}_{\ce{H_2}}) \le$ \num{-30} at the inlet of the domain correspond to flux magnitudes of \SI{1e-30}{\kilo\gram\per\square\meter\per\second} or less.
At the far right of Figures~\ref{H2_contours (g)} and~\ref{H2_contours (h)}, the relative angles for both models are roughly constant at $\pi$, anti-parallel to the species gradient vector.
In this region fluxes are small but non-zero as residual fuel is present in small concentrations---as a result, scalar gradients are small.
Finally, although the flux angle appears to deviate from $\pi$ in small regions throughout the flame for both the mixture-averaged and multicomponent flames, these deviations correspond to areas where the flux magnitude is locally very small, approaching zero.
Furthermore, the relative angle of the flux vector is consistently $\pi$ in regions of high species gradients, such as through the flame front, and agrees well between the two models.
The angles between the mass-flux and gradient vectors agree similarly well for all species and flame configurations considered. 

% \begin{figure}[htbp]
% \centering
%     \begin{subfigure}{0.48\textwidth}
%         \centering
%             \includegraphics[width=\textwidth]{Angle_PDF_2D.pdf}
%         \caption{Two-dimensional}\label{PDF (a)}
%     \end{subfigure}
%     ~
%     \begin{subfigure}{0.48\textwidth}
%         \centering
%             \includegraphics[width=\textwidth]{Angle_PDF_3D.pdf}
%         \caption{Three-dimensional}\label{PDF (b)}
%     \end{subfigure}
%     \caption{A priori assessment of the mixture-averaged and multicomponent models, comparing the PDFs of the angle between species flux vectors and species gradient vectors for the two-dimensional unsteady (\subref{PDF (a)}) and three-dimensional turbulent (\subref{PDF (b)}) hydrogen/air flames.
%     The inset plots use a semi-log scale on the vertical axis.}
%     \label{PDF}
% \end{figure}

\begin{figure}[htbp]
\centering
\includegraphics[width=\textwidth]{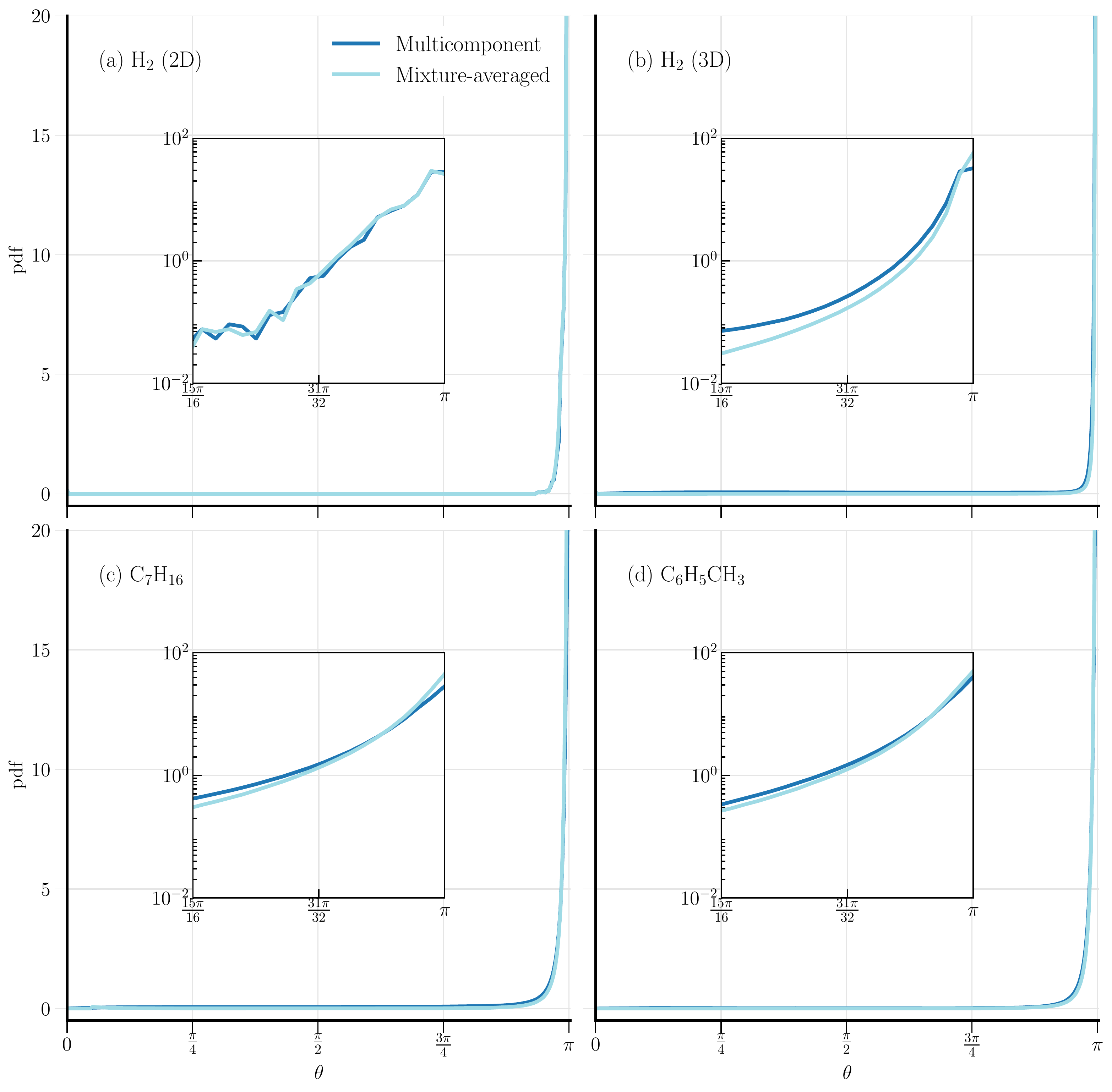}
\caption{A priori assessment of the mixture-averaged and multicomponent models, comparing the PDFs of the angle between species flux vectors and species gradient vectors: $\theta = \nabla Y_F \angle \mathbf{j}_{F}$, where $F$ is the fuel in the (a) two-dimensional unsteady hydrogen, (b) three-dimensional turbulent hydrogen, (c) three-dimensional turbulent \textit{n}-heptane, and (d) three-dimensional turbulent toluene flames.
The inset plots use a semi-log scale on the vertical axis.}
\label{fig:angle_pdfs}
\end{figure}

To confirm our qualitative observations of the relative direction of the flux vector, Figure~\ref{fig:angle_pdfs} shows the probability density function (PDF) of the angles between the fuel species diffusion flux vector and mass fraction gradient vector for the two-dimensional unsteady hydrogen flame, three-dimensional turbulent hydrogen flame, turbulent \textit{n}-heptane flame, and turbulent toluene flame.
This quantitatively measures the alignment of the vectors to compare the multicomponent and mixture-averaged models.
We only consider points in the domain where the species diffusion flux magnitude is at least \SI{10}{\percent} of the peak species diffusion flux magnitude, to emphasize the regions where diffusion is important.

In the two- and three-dimensional hydrogen flames, as shown in Figures~\ref{fig:angle_pdfs}(a) and \ref{fig:angle_pdfs}(b), both the mixture-averaged and multicomponent diffusion models have maximum PDF values at an angle of $\pi$, anti-parallel to the species gradient vector. 
As expected, this indicates that mass diffusion occurs primarily in the direction of negative species gradient (i.e., from high to low concentration).
We attribute small deviations of the mixture-averaged angle away from $\pi$ to the velocity correction term in Eq.~\eqref{eq:MA-diffusion-flux}.

The two-dimensional unsteady flame exhibits negligible differences in the angles separating the species diffusion flux and gradient vectors,
but this agreement does not extend to the three-dimensional turbulent flame.
In this case, the angle PDF is roughly \SI{50}{\percent} higher for the multicomponent model.
Although this difference between the two models is large, it is tempered by the tiny magnitude of the PDFs away from $\pi$.
For both cases these vectors show a clear preferential alignment at $\pi$, with the magnitude of the PDF dropping to much less than one for angles smaller than $63\pi/64$.
As seen in Figures~\ref{fig:angle_pdfs}(c) and \ref{fig:angle_pdfs}(d), the three-dimensional turbulent \textit{n}-heptane/air and toluene/air flames show similar differences.

% \begin{figure}[htbp]
% \centering
%     \begin{subfigure}{0.48\textwidth}
%         \centering
%         \includegraphics[width=\textwidth]{Hep_Angle_PDF_3D.pdf}
%         \caption{\ce{n-C7H16}}
%         \label{fig:pdf_heptane}
%     \end{subfigure}
%     ~
%     \begin{subfigure}{0.48\textwidth}
%         \centering
%         \includegraphics[width=\textwidth]{Tol_Angle_PDF_3D.pdf}
%         \caption{\ce{C6H5CH3}}
%         \label{fig:pdf_toluene}
%     \end{subfigure}
%      \caption{A priori assessment of the mixture-averaged and multicomponent models, comparing the PDFs of the angle between species flux vectors and species gradient vectors for the three-dimensional turbulent \textit{n}-heptane/air (\subref{fig:pdf_heptane}) and toluene/air (\subref{fig:pdf_toluene}) flames. 
%      The inset plots are semi-log scale on the $y$ axis.}
%      \label{PDF_2}
% \end{figure}

% It is plausible that increases in local flame curvature, characteristic of highly turbulent flames, may reduce the accuracy of the mixture-averaged model.
% In this two-dimensional configuration the orientation of the mixture-averaged diffusion flux provides an excellent approximation to the orientation of the multicomponent model over the entire range of data.

\begin{figure}[htbp]
\centering
    \begin{subfigure}{0.65\textwidth}
        \centering
            \includegraphics[width=\textwidth]{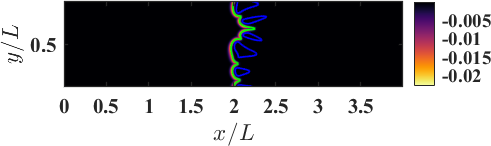}
        \caption{Two-dimensional \ce{H2}}
        \label{flux_diff_contours (a)}
    \end{subfigure}\\
    \begin{subfigure}{0.65\textwidth}
        \centering
            \includegraphics[width=\textwidth]{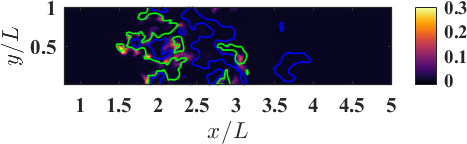}
        \caption{Three-dimensional \ce{H2}}
        \label{flux_diff_contours (b)}
    \end{subfigure}\\
    \begin{subfigure}{0.65\textwidth}
        \centering
            \includegraphics[width=\textwidth]{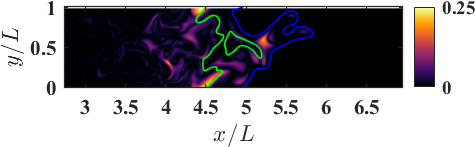}
        \caption{\ce{n-C7H16}}
        \label{flux_diff_contours (c)}
    \end{subfigure}\\
    \begin{subfigure}{0.65\textwidth}
        \centering
            \includegraphics[width=\textwidth]{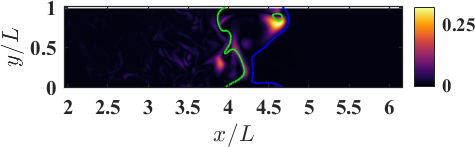}
        \caption{\ce{C6H5CH3}}
        \label{flux_diff_contours (d)}
    \end{subfigure}
  \caption{Local differences between mixture-averaged and multicomponent mass diffusion fluxes of the fuel normalized by peak multicomponent mass diffusion flux: $\left( \left| \textbf{j}^{\text{MA}} \right| - \left| \textbf{j}^{\text{MC}} \right| \right) \big/ \max\limits_{N_p} \left(\left| \textbf{j}^{\text{MC}} \right|\right)$. 
  Shown are domain cross-sections through the midplane. 
  The green and blue lines correspond to isosurfaces of $T_u=T_{\text{peak}}-\SI{300}{\kelvin}$ and $T_b=T_{\text{peak}}+\SI{300}{\kelvin}$, respectively, and represent the inflow and outflow surfaces of the flame front.}
  \label{flux_diff_contours}
\end{figure}

To better show where the two models deviate in the domain, Figure~\ref{flux_diff_contours} presents contours of the point-wise difference between the fuel diffusion flux magnitudes between the models, normalized by the peak multicomponent fuel diffusion-flux magnitude.
These contour plots provide a reference for the physical location of peak differences between the mixture-averaged and multicomponent models within the flame.

Examining Figure~\ref{flux_diff_contours}, in all cases the highest differences between the two models occur in regions of high flame curvature where the species gradient field is strong and highly variable.
To more-concisely discuss these differences across multiple species and flame configurations, Table~\ref{tab:L2_norms} presents the mean and standard deviations of the angles between the mixture-averaged and multicomponent diffusion fluxes, as well as relative $L^1$, $L^2$, and $L^{\infty}$ error norms of the differences in magnitude of these diffusion fluxes.
We calculated these statistics in regions where species diffusion is strong, i.e., the diffusion flux magnitude is greater than \SI{10}{\percent} of the peak diffusion flux magnitude.
The relative $L^1$, $L^2$, and $L^{\infty}$ error norms measure the modal, mean, and maximum difference, respectively, for the diffusion flux magnitude:
{\allowdisplaybreaks \begin{IEEEeqnarray}{rCl}
L^1\left(\mathbf{j}_i^{\text{MA}}\right) &=& \frac{\sum_{n=1}^{N_p}\left| \left| \textbf{j}_{i,n}^{\text{MA}} \right| - \left| \textbf{j}_{i,n}^{\text{MC}} \right| \right|}{\sum_{n=1}^{N_p} \left| \textbf{j}_{i,n}^{\text{MC}} \right|} \;,
\label{eq:L1} \\
L^2\left(\mathbf{j}_i^{\text{MA}}\right) &=& \sqrt{\frac{\sum_{n=1}^{N_p} \left( \left| \textbf{j}_{i,n}^{\text{MA}} \right| - \left| \textbf{j}_{i,n}^{\text{MC}} \right| \right)^{2}}{\sum_{n=1}^{N_p} \left| \textbf{j}_{i,n}^{\text{MC}} \right|^{2}}} \;, \text{ and}
\label{eq:L2} \\
L^\infty\left(\mathbf{j}_i^{\text{MA}}\right) &=& \frac{\max\limits_{n \in N_p} \left( \left| \textbf{j}_{i,n}^{\text{MA}} \right| - \left| \textbf{j}_{i,n}^{\text{MC}} \right| \right)}{\max\limits_{n \in N_p} \left(\left| \textbf{j}_{i,n}^{\text{MC}} \right|\right)} \;,
\label{eq:Linfinity}
\end{IEEEeqnarray}}%
where $N_p$ is the number of points in the domain where the diffusion flux is greater than 10\% of the peak magnitude and $\textbf{j}_{i,n}$ indicates the diffusion flux of the $i$th species at point $n$.
% The relative L2 error represents the accuracy in the mixture-averaged diffusion flux relative to the multicomponent flux.
%A similar relative $L^2$ error norm can be defined for the angles between species diffusion fluxes for the two diffusion models.

\begin{table}[htbp]
\scriptsize
    \centering
    \caption{Statistical quantities of the mixture-averaged and multicomponent diffusion models for a representative set of major, radical, and product species: the mean ($\mu_{\angle}$) and standard deviation ($s_{\angle}$) of the angles between the mixture-averaged and multicomponent flux vectors, as well as relative $L^2$ error norms (Eq.~\eqref{eq:L2}).}
    %\begin{tabular}{@{\extracolsep{\fill}}l c c c@{}}
    \begin{tabular}{@{}l c c c c c@{}}
         \toprule
         & $\mu_{\angle}$ [\si{\radian}] & $s_{\angle}$ [\si{\radian}] & $L^1\left(\mathbf{j}_i^{\text{MA}}\right)$ & $L^2\left(\mathbf{j}_i^{\text{MA}}\right)$ & $L^\infty\left(\mathbf{j}_i^{\text{MA}}\right)$ \\
         \midrule
         \multicolumn{6}{c}{2D unsteady hydrogen} \\
         \midrule
         \ce{H2} & \num{2.7e-5} & \num{2.1e-4} & \num{2.8e-2} & \num{2.6e-2} & \num{2.2e-2} \\
         \ce{H} & \num{5.7e-5} & \num{1.4e-6} & \num{3.8e-2} & \num{3.5e-2} & \num{2.8e-2} \\
         \ce{OH} & \num{2.4e-4} & \num{3.2e-5} & \num{1.2e-2} & \num{1.0e-2} & \num{8.7e-3} \\ 
         \ce{H2O} & \num{7.4e-3} & \num{2.1e-4} & \num{4.2e-2} & \num{4.0e-2} & \num{4.1e-2} \\ 
         \midrule
         \multicolumn{6}{c}{3D hydrogen} \\
         \midrule
         \ce{H2} & \num{3.3e-2} & \num{7.3e-2} & \num{1.5e-1} & \num{2.1e-1} & \num{3.4e-1} \\
         \ce{H} & \num{2.5e-2} & \num{5.1e-3} & \num{1.4e-1} & \num{2.1e-1} & \num{4.2e-1} \\
         \ce{OH} & \num{7.1e-4} & \num{1.4e-4} & \num{1.6e-1} & \num{2.2e-1} & \num{4.0e-1} \\
         \ce{H2O} & \num{1.5e-1} & \num{7.3e-3} & \num{1.8e-1} & \num{2.5e-1} & \num{4.6e-1} \\
         \midrule
         \multicolumn{6}{c}{3D \textit{n}-heptane} \\
         \midrule
         \textit{n}-\ce{C7H16} & \num{4.8e-2} & \num{4.7e-2} & \num{1.7e-1} & \num{2.2e-1} & \num{4.1e-1} \\
         \ce{H} & \num{2.7e-2} & \num{5.7e-2} & \num{1.5e-1} & \num{2.0e-1} & \num{3.9e-1} \\
         \ce{OH} & \num{6.0e-2} & \num{5.9e-2} & \num{1.7e-1} & \num{2.3e-1} & \num{4.1e-1} \\
         \ce{CO2} & \num{4.8e-3} & \num{1.3e-3} & \num{1.5e-1} & \num{2.0e-1} & \num{3.3e-1} \\
         \midrule
         \multicolumn{6}{c}{3D toluene} \\
         \midrule
         \ce{C6H5CH3} & \num{6.1e-2} & \num{5.1e-2} & \num{2.2e-1} & \num{2.8e-1} & \num{4.7e-1} \\
         \ce{H} & \num{1.4e-2} & \num{4.8e-2} & \num{1.8e-1} & \num{2.5e-1} & \num{4.0e-1} \\
         \ce{OH} & \num{8.6e-3} & \num{2.7e-2} & \num{2.1e-1} & \num{2.7e-1} & \num{4.6e-1} \\
         \ce{CO2} & \num{5.9e-3} & \num{2.6e-3} & \num{2.0e-1} & \num{2.5e-1} & \num{3.9e-1} \\
         \bottomrule
    \end{tabular}
    \label{tab:L2_norms}
\end{table}

As observed in Table~\ref{tab:L2_norms}, a majority of the mixture-averaged diffusion flux vectors match the multicomponent diffusion flux vectors within a mean angle, $\mu_{\angle}$, of \SI{0.06}{\radian} for the turbulent cases, with negligible differences for the two-dimensional case.
The exception is the diffusion flux of \ce{H2O} for the hydrogen/air turbulent flame with a mean angle of \SI{0.12}{\radian}.
As expected, the turbulent cases show larger (albeit still small) values of $\mu_{\angle}$.
%The effects of turbulence reduce the agreement between the mixture-averaged and multicomponent diffusion flux angles, yet they still remain within $\sim$\SI{0.1}{\radian}.
Additionally, the standard deviations of the angle between the diffusion fluxes, $s_{\angle}$, are small and generally the same order as the mean angles themselves.%, the angles will statistically remain small ($\sim$\SI{0.1}{\radian}) and close to zero.}}

Finally, Table~\ref{tab:L2_norms} shows that the magnitudes of the diffusion fluxes agree throughout much of the domain.
The $L^2$ error norms indicate the average difference in the mixture-averaged flux magnitude is on the order of \SI{20}{\percent}.
However, by definition, the $L^2$ norm is sensitive to outliers, which increase these errors when present. 
Alternatively, the $L^1$ error norm weights all points in the domain equally, providing a measure of the modal error.
Comparing these two values, we can see the differences are smaller than \SI{20}{\percent} throughout the domain for most species, with the largest differences occurring in the three-dimensional flame configurations.
Finally, examining the $L^\infty$ error norms, the observed differences in Figure~\ref{flux_diff_contours} correspond to large differences, on the order of \SIrange{30}{50}{\percent}, in regions of high flame curvature for each of the three-dimensional flame configurations.
The largest differences occur in the \textit{n}-heptane/air and toluene/air flames.

The differences between the two models seem to increase proportionally to the magnitude of the driving species gradient.
In Figure~\ref{flux_diff_contours (a)}, showing the lean, two-dimensional, unsteady, laminar, hydrogen/air flame, the mixture-averaged model matches the multicomponent model within \SI{2}{\percent} for the full domain.
For this two-dimensional case the species gradient vectors are primarily aligned in the flow-wise direction and roughly constant across the domain.
As a result, the scalar gradient fields locally vary a small amount and the mixture-averaged diffusion model matches the multicomponent model well; 
this is true even near thermal instabilities.
In contrast, the lean, three-dimensional, turbulent hydrogen air flame shows significantly larger differences between the models.
The equivalence ratio between these two flames matches ($\phi = 0.4$) and so any local increases in the species gradient field come from increases in local flame curvature caused by a higher dimensionality and turbulent mixing.
In these regions of high-flame curvature, the scalar gradient field is steep, highly variable, and not strictly aligned in the flow-wise direction; as a result, 
mass can diffuse in directions other than the direction of the species gradient.
Comparing the definitions of the mixture-averaged and multicomponent diffusion fluxes in Eq.~\eqref{eq:MA-diffusion-flux} and Eq.~\eqref{eq:MC-diffusion-flux}, respectively, the strict alignment of the mixture-averaged diffusion flux with its own gradient may overvalue the impact of that gradient and overpredict the mass flux when the gradient vector is large, such as in regions of high flame curvature.

This overprediction in the mixture-averaged diffusion flux is most evident in the turbulent, \textit{n}-heptane/air and toluene/air flames where, in addition to turbulent mixing, the equivalence ratio is higher than in the turbulent hydrogen\slash air flame ($\phi=0.9$ compared with $\phi=0.4$).
Although it is difficult to compare these flames one-to-one due to differences in chemistry, the increased equivalence ratio relative to the lean hydrogen/air flames causes steeper species gradients through the flame, since the unburnt fuel mass fractions are much higher in these flames: \num{0.056} and \num{0.063} for \textit{n}-heptane/air and toluene/air, respectively, compared with \num{0.012} for hydrogen/air.
This increase in the unburnt fuel mass fraction, coupled with the presence of turbulent mixing, increases the magnitude of the species gradients through the flame front relative to the three-dimensional hydrogen flame.
Furthermore, the larger differences in the \textit{n}-heptane and toluene flames over the hydrogen/air flames supports the theory that high species gradients can lead to the mixture-averaged model overpredicting diffusion flux.

Finally, although the reported differences in the predicted flux magnitudes for the mixture-averaged over the multicomponent model may seem significant, it is important to examine these values from the perspective of turbulent scaling.
The relative magnitude of both the mixture-averaged and multicomponent mass diffusion fluxes are small: on the order of \SI{5e-3}{\kilo\gram\per\square\meter\per\second} or smaller on average for all four simulations, compared to total mass fluxes on the order of \SI{1}{\kilo\gram\per\square\meter\per\second} or greater.
Moreover, the multicomponent and mixture-averaged fluxes consistently show the same order of magnitude for a given species in a point-wise comparison, for most points in the domain.
In other words, when comparing any given point in the domain, the expected fluxes have the same order of magnitude for the mixture-averaged and multicomponent models.
This suggests that the differences in the mixture-averaged flux may be small compared to global turbulent flame statistics.

\subsection{A posteriori comparison of turbulent statistics}\label{sec:aposterioriresults}
To further compare the mixture-averaged and multicomponent diffusion models, we present a posteriori statistics for the three turbulent flame simulations using both mixture-averaged and multicomponent diffusion.
In other words, how much does model choice impact global and statistical quantities in a simulation?
For this analysis, we started simulations using the multicomponent and mixture-averaged diffusion models from the same initial conditions, and allowed them to evolve in the domain until any initial transients had convected through the domain: approximately six eddy turnover times ($\tau_{\text{eddy}} = k/\epsilon$, where $k$ is the turbulent kinetic energy)\footnote{In practice we determine $\tau_{\text{eddy}}$ with the turbulence forcing scheme:  $\tau_{\text{eddy}}=1/2A_{\text{force}}$, where $A_{\text{force}}$ is the turbulent forcing coefficient imposed as a parameter in the simulation, developed by Carroll et al.~\cite{Carroll2013}.}.
We then ran each simulation for an additional $20 \tau_{\text{eddy}}$ to provide a representative sample of instantaneous flame speeds and collect turbulent statistics.
% we allowed the flames to develop in a turbulent flowfield, and computed statistics after removing the initial transients of the input flow and scalar fields.
%We ran each simulation for at least 20 eddy turnover times ($\tau_{\text{eddy}} = k/\epsilon$, where $k$ is the turbulent kinetic energy) to collect turbulent statistics\footnote{In practice we determine $\tau_{\text{eddy}}$ with the turbulence forcing scheme:  $\tau_{\text{eddy}}=1/2A_{\text{force}}$, where $A_{\text{force}}$ is the turbulent forcing coefficient imposed as a parameter in the simulation, developed by Carroll et al.~\cite{Carroll2013}.}.

\begin{figure}[htbp]
\centering
    % \begin{subfigure}{0.5\textwidth}
    %     \centering
    %     \includegraphics[width=\textwidth]{H2_Flame_history.pdf}
    %     \caption{\ce{H2}}
    % \end{subfigure}\\
    % \begin{subfigure}{0.5\textwidth}
    %     \centering
    %     \includegraphics[width=\textwidth]{Hep_Flame_history.pdf}
    %     \caption{\ce{C7H16}}
    % \end{subfigure}\\
    % \begin{subfigure}{0.5\textwidth}
    %     \centering
    %     \includegraphics[width=\textwidth]{Tol_Flame_history.pdf}
    %     \caption{\ce{C6H5CH3}}
    % \end{subfigure}
\includegraphics[width=0.9\textwidth]{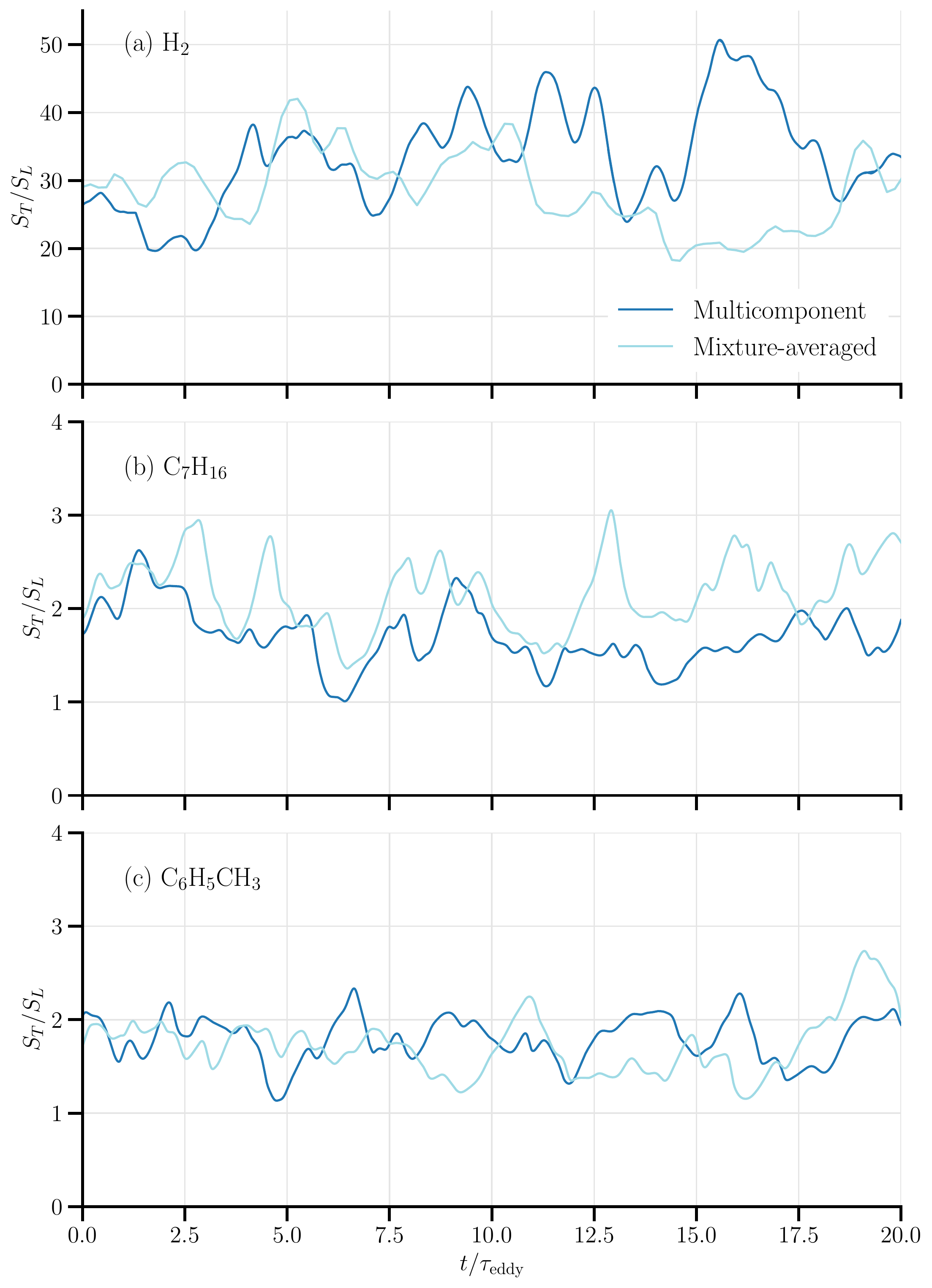}
\caption{Time histories of the normalized turbulent flame speed from the turbulent (a) hydrogen\slash air, (b) \textit{n}-heptane\slash air, and (c) toluene\slash air cases for both diffusion models.}
\label{Flame speed}
\end{figure}

Figure~\ref{Flame speed} shows the time-history of the turbulent flame speed, $S_T$, normalized by the laminar flame speed, $S_L$.
The turbulent flame speed is defined as
\begin{equation}
S_{T} = - \frac{\int_{V} \rho \dot{\omega}_{\ce{F}} \, dV}{\rho_{u} Y_{\ce{F},u} L} \;,
\end{equation}
where $\dot{\omega}_{\ce{F}}$ and $Y_{F}$ are the fuel source term and mass fraction respectively, $\rho$ is the density, $L$ is the span-wise domain width, and $V$ is the volume of the domain.
Table~\ref{tab:flame_speeds} presents the mean turbulent flame speeds of the three fuels, based on the collected samples;
we report these values to provide a simple, single metric to identify global differences between the models.
While the mixture-averaged model seems to lower the normalized turbulent flame speeds by \SI{13}{\percent} and \SI{5}{\percent} for the hydrogen and toluene flames, it causes a \SI{20}{\percent} higher normalized flame speed for the \textit{n}-heptane flame.
These trends do not seem to correlate to the Lewis number for each fuel---recall that $\text{Le}_{\ce{H2}} = 0.3$, $\text{Le}_{\ce{C6H5CH3}} = 2.5$, and $\text{Le}_{\ce{C7H16}} = 2.8$.
The \textit{n}-heptane/air and toluene/air flames have similar Lewis numbers but show opposite trends in the differences of turbulent flame speeds between the diffusion models.
However, the differences are small between the models for toluene, and may be attributed to statistical error.
In contrast, the differences between the models are larger for the hydrogen and \textit{n}-heptane flames (13\% and 20\%, respectively); 
these values are similar to the spread in mean turbulent flame speed ($\sim$15\%) for \textit{n}-heptane flames with varying equivalence ratio and chemical model shown by Lapointe and Blanquart~\cite{Lapointe2016FuelFlames}.

\begin{table}[htbp]
\scriptsize
    \centering
    \caption{Mean turbulent flame speed normalized by unstretched laminar flame speed ($S_T/S_L$) for three-dimensional turbulent hydrogen/air, \textit{n}-heptane/air, and toluene/air mixtures, comparing the impact of mixture-averaged and multicomponent diffusion models.}
    \begin{tabular}{@{}l c c c@{}}
        \toprule
        & MC & MA & Difference\\
        \midrule
        \ce{H2} & \num{33.9} & \num{29.6} & \SI{13}{\percent} \\ 
        \ce{C7H16} & \num{1.70} & \num{2.02} & \SI{20}{\percent} \\
        \ce{C6H5CH3} & \num{1.80} & \num{1.71} & \SI{5}{\percent}\\
        \bottomrule
    \end{tabular}
    \label{tab:flame_speeds}
\end{table}

To better understand the observed differences in the turbulent flame speed, we assess the impact of the diffusion models on flame chemistry via the average fuel mass fraction and source term.
As demonstrated in previous studies, differential diffusion can modify the local equivalence ratio in regions of high flame curvature~\cite{Lapointe:2015,Lapointe2016FuelFlames,peters2001turbulent}.
We have demonstrated this effect in Table~\ref{tab:L2_norms}, by showing that the mixture-averaged diffusion assumption overpredicts the magnitude of the mass diffusion flux in these regions.
The increase in mass flux into these regions of high flame curvature may impact local chemistry and modify the fuel source term.
Lapointe and Blanquart~\cite{Lapointe2016FuelFlames} previously suggested that the normalized turbulent flame speed is proportional to the product of turbulent flame area and the mean fuel source term conditioned on flame temperature:
\begin{equation}\label{LaPoint_Blanquart}
    \frac{S_T}{S_L}\propto\frac{A_T}{A} \frac{\langle\dot{\omega}_{F}|T\rangle}{\dot{\omega}_{F,\text{lam}}} \;,
\end{equation}
where $A_T$ is the turbulent flame area, $A$ is the cross-sectional area of the domain, and $\dot{\omega}_{F,\text{lam}}$ is the fuel source term in the laminar flame.
Moreover, they demonstrated that the area ratio ($A_T/A$) controls large-scale fluctuations in the flame speed on the order of their mean values, while the normalized mean source term ($\langle\dot{\omega}_{F}|T\rangle / \dot{\omega}_{F,\text{lam}}$) controls smaller-scale fluctuations~\cite{Lapointe2016FuelFlames}.

\begin{figure}[htbp]
  \centering
  \includegraphics[width=\textwidth]{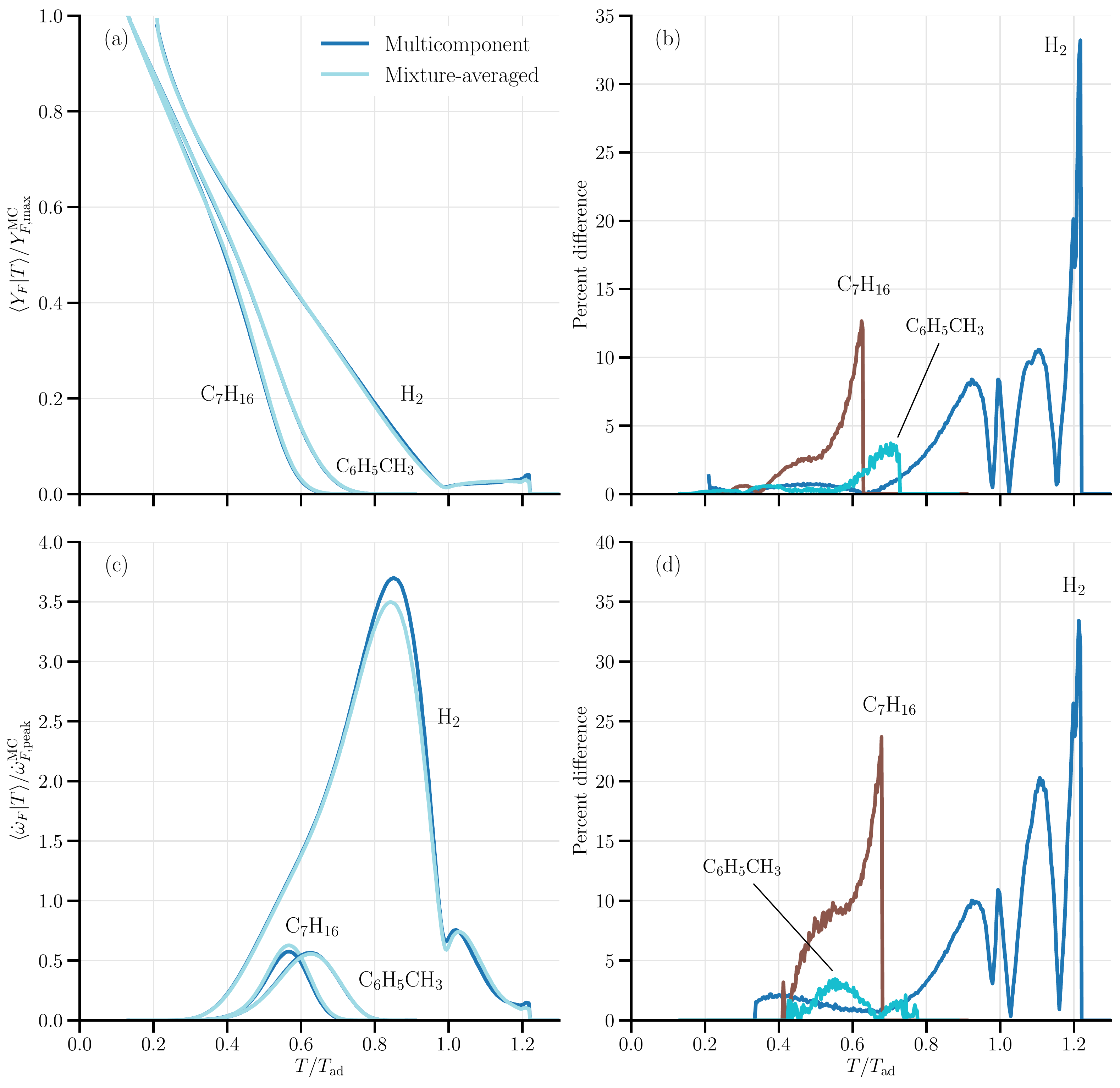}
  \caption{Turbulent flame structure for the three fuel/air mixtures, showing conditional means (left) and percent differences (right) of the mixture-averaged values with respect to the multicomponent values, for (a--b) fuel mass fraction and (c--d) fuel source term, as functions of temperature $T$ normalized by $T_{\text{ad}}$. 
  All values are normalized by their peaks from one-dimensional flat flames using the multicomponent model.}
  \label{fig:conditional_means}
\end{figure}

To evaluate if the observed difference in the diffusion mass fluxes correlate to the observed differences in the normalized turbulent flame speeds between the two models, Figures~\ref{fig:conditional_means} (a) and (c) present the means of the fuel mass fractions and source term, conditioned on temperature and shown normalized by their respective adiabatic flame temperatures, $T_{\text{ad}}$.
Figures~\ref{fig:conditional_means} (b) and (d) show the percent differences between the conditional means of fuel mass fraction and source term, respectively, with respect to the multicomponent model results;
only differences corresponding to normalized conditional means from the multicomponent model above 0.01 and 0.05 are shown, respectively.
The calculated conditional means of the fuel mass fractions differ by negligible amounts for most of the domain: less than \SI{1}{\percent} for most points, with only larger differences (i.e., greater than \SI{5}{\percent}) between small values of the mean fuel mass fraction.
This strong agreement between the conditional means of the fuel mass fraction suggests that although local differences in the diffusion mass flux fields do exist between the two models, they do not appear to significantly affect the averaged distribution of fuel in the flame.

In contrast, we observe more differences in the conditional means of the fuel source term in Figures~\ref{fig:conditional_means} (c) and (d).
At their peaks, the mean source terms for the multicomponent flames are \SI{5.5}{\percent} and \SI{1.6}{\percent} higher than the mixture-averaged models for the hydrogen and toluene flames, respectively, and \SI{9.4}{\percent} lower for the \textit{n}-heptane flame.
Away from the peak locations, the conditional means disagree by about \SIrange{1}{35}{\percent} for the hydrogen\slash air flame, and by up to about \SI{24}{\percent} for the \textit{n}-heptane\slash air flame.
However, most differences occur where the values of mean source term are small.
For the toluene\slash air flame, the mixture-averaged and multicomponent cases agree within \SI{3.5}{\percent} at all locations.

For the hydrogen\slash air flame, we see differences of \SIrange{5}{35}{\percent} between the mixture-averaged and multicomponent in the super-adiabatic regions, for the conditional means of both the fuel mass fraction and source term.
These regions, also called ``hot spots'',  result from differential diffusion, and have been predicted both in theoretical studies \cite{Williams:1985} and in numerical analyses of lean hydrogen/air mixtures~\cite{Day:2009,AspdenJFM:2011,Aspden:2015}.
%This agreement also extends into super-adiabatic regions for the hydrogen/air flame.
%

Comparing the observed differences in the conditional means of the fuel source term with the mean normalized turbulent flame speeds, the results agree well with the proportional relation in Eq.~\eqref{LaPoint_Blanquart} given by Lapointe and Blanquart~\cite{Lapointe2016FuelFlames}.
Specifically, the peak normalized source term is \SI{5.5}{\percent} higher for the multicomponent hydrogen/air flame over mixture-averaged, resulting in a \SI{13}{\percent} higher normalized flame speed.
Alternatively, the peak normalized source term is \SI{9.4}{\percent} lower for the multicomponent n-heptane/air flame compared to mixture-averaged, resulting in a \SI{19}{\percent} lower normalized flame speed.
In all three cases, the differences in normalized turbulent flame speed appear proportional to the differences in normalized conditional mean by approximately a factor of two.
This strong proportional relationship agrees with similar results observed by Lapointe and Blanquart~\cite{Lapointe2016FuelFlames}.
Moreover, our results demonstrate that, relative to the multicomponent model, the observed differences in the mixture-averaged diffusion flux vectors do impact a posteriori flame statistics.
These results raise questions about the appropriateness of the mixture-averaged diffusion assumption for simulations with high flame curvature.

\subsection{Comparison of chemical pathways}
\label{sec:pathways}

Mass diffusion plays an important role in premixed flames, since diffusion of radical species can alter elementary reaction rates in the reaction zone.
In this section, we examine how the choice of diffusion model impacts the relative contribution of major reactions to fuel consumption.

We performed additional one-dimensional, unstretched (flat), laminar flame simulations using the \texttt{FreeFlame} Cantera v2.5.0~\cite{cantera} for each of the three fuel\slash air mixtures, corresponding to the unburnt conditions given in Table~\ref{tab:2.3D_flow_config}.
We used the freely-propagating adiabatic ﬂat ﬂame solver (\texttt{FreeFlame})  with grid reﬁnement criteria for both slope and curvature set to 0.01 and a reﬁnement ratio of 2.0.

In the laminar simulations, these reactions account for more than \SI{98}{\percent} of the overall fuel consumption rate in the hydrogen flame:
\begin{align}
    \ce{H2 + O &<-> H + OH} \\
    \ce{H2 + OH &<-> H + H2O} \;,
\end{align}
in the \textit{n}-heptane flame:
\begin{align}
    \ce{$n${-}C7H16 + H &-> 2-C7H15 + H2} \\
    \ce{$n${-}C7H16 + O &-> 2-C7H15 + OH} \\
    \ce{$n${-}C7H16 + OH &-> 2-C7H15 + H2O} \;,
\end{align}
and in the toluene flame:
\begin{align}
    \ce{C6H5CH3 + H &<-> C6H6 + CH3} \\
    \ce{C6H5CH3 + H &-> C6H5CH2 + H2} \label{rxn:toluene1} \\
    \ce{C6H5CH3 + OH &-> C6H5CH2 + H2O} \label{rxn:toluene2} \;.
\end{align}%
Tables~\ref{tab:hydrogen_pathways}, \ref{tab:heptane_pathways}, and \ref{tab:toluene_pathways} list the percentage contributions of the primary reactions to the overall fuel consumption rate of the corresponding hydrogen, \textit{n}-heptane, and toluene flames, respectively.
In the hydrogen and toluene flames, we focus on the reactions that contribute most to fuel consumption only, and leave out reactions that re-form the fuel.

\begin{table}[htbp]
\centering
\caption{Percentage contribution of major reactions to the overall fuel consumption rate in the hydrogen flames. 
``MC'' and ``MA'' represent cases using the multicomponent and mixture-averaged diffusion models, respectively.}
\begin{tabular}{@{}l c c c@{}}
    \toprule
    Reaction & Model & 1D & Turbulent \\
    \midrule
    \ce{H2 + O <-> H + OH} & MC & 11.5 & 17.5 \\
      & MA & 11.5 & 17.7 \\
    \rule{0pt}{4ex} \ce{H2 + OH <-> H + H2O} & MC & 86.5 & 82.5 \\
      & MA & 86.5 & 82.3 \\
    \bottomrule
\end{tabular}
\label{tab:hydrogen_pathways}
\end{table}

\begin{table}[htbp]
\centering
\caption{Percentage contribution of major reactions to the overall fuel consumption rate of the \textit{n}-heptane flames. 
``MC'' and ``MA'' represent cases using the multicomponent and mixture-averaged diffusion models, respectively.}
\begin{tabular}{@{}l c c c@{}}
    \toprule
    Reaction & Model & 1D & Turbulent \\
    \midrule
    \ce{$n${-}C7H16 + H -> 2-C7H15 + H2} & MC & 59.5 & 53.5 \\
      & MA & 59.3 & 55.4 \\
    \rule{0pt}{4ex} \ce{$n${-}C7H16 + O -> 2-C7H15 + OH} & MC & 7.78 & 8.80 \\
      & MA & 7.83 & 8.95 \\
    \rule{0pt}{4ex} \ce{$n${-}C7H16 + OH -> 2-C7H15 + H2O} & MC & 31.7 & 37.0 \\
      & MA & 31.8 & 34.7 \\
    \bottomrule
\end{tabular}
\label{tab:heptane_pathways}
\end{table}

\begin{table}[htbp]
\centering
\caption{Percentage contribution of major reactions to the overall fuel consumption rate of the toluene flames, neglecting fuel re-formation reactions.
``MC'' and ``MA'' represent cases using the multicomponent and mixture-averaged diffusion models, respectively.}
\begin{tabular}{@{}l c c c@{}}
    \toprule
    Reaction & Model & 1D & Turbulent \\
    \midrule
    \ce{C6H5CH3 + H <-> C6H6 + CH3} & MC & 22.4 & 20.5 \\
      & MA & 22.4 & 20.7 \\
    \rule{0pt}{4ex} \ce{C6H5CH3 + H -> C6H5CH2 + H2} & MC & 42.1 & 36.3 \\
      & MA & 42.0 & 36.1 \\
    \rule{0pt}{4ex} \ce{C6H5CH3 + OH -> C6H5CH2 + H2O} & MC & 35.1 & 43.0 \\
      & MA & 35.2 & 43.0 \\
    \bottomrule
\end{tabular}
\label{tab:toluene_pathways}
\end{table}

These results show that the choice of diffusion model does not significantly modify the primary reaction pathways for fuel consumption.
In the one-dimensional laminar flames, the percentage contributions of each reaction are nearly identical for all fuels (i.e., within \SI{0.1}{\percent}).
In the turbulent flames, the \textit{n}-heptane flame shows differences of about \SI{2}{\percent} between the diffusion models, with even smaller differences in the hydrogen and toluene flames.
Overall, the choice of diffusion model impacts reaction pathways less than turbulence.
In fact, in the case of toluene, turbulence changes the ordering of primary reactions; both diffusion models equally capture these effects.
Our results agree with and extend those of Lapointe et al.~\cite{Lapointe:2015}, who previously showed that using two Lewis-number models (unity and non-unity Lewis number) did not substantially change the primary reaction pathways in high-Karlovitz turbulent premixed \textit{n}-heptane\slash air flames, like those studied here.

\section{Conclusions}
\label{sec:conclusions}

This article compares the mixture-averaged and multicomponent mass diffusion models for premixed two-dimensional, unsteady hydrogen/air and three-dimensional, turbulent flames, considering hydrogen, \textit{n}-heptane, and toluene fuel/air mixtures.
We compared the methods using both a priori and a posteriori assessments of differences.

The a priori analysis indicated that the mixture-averaged model accurately reproduces the relative direction and magnitude of the flux vectors through much of the domain.
However, in the turbulent cases, we found average differences of \SIrange{10}{20}{\percent} in the magnitude of the diffusion flux vector for all three fuels, 
and differences greater than \SI{40}{\percent} in regions of high flame curvature.

Our a posteriori analysis indicated that using the mixture-averaged model does affect turbulent statistics, such as conditional means of the fuel mass fraction and consumption rates.
The impact on flame statistics is relatively small: the mixture-averaged model results in differences of up to \SI{20}{\percent} in normalized turbulent flame speed, \SI{10}{\percent} in conditional mean of fuel source term, and \SI{1}{\percent} in the conditional mean of fuel mass fraction.
Thus, the larger differences we observed in instantaneous diffusion fluxes seem to lead to smaller---though nonzero---differences in average quantities.
The differences between the diffusion models also do not significantly impact the relative contributions of reactions to fuel consumption.
Due to differences between the three turbulent flames examined, notably the varying Karlovitz numbers and flame stability, we are unable to draw firm conclusions on the root causes of the observed differences.
These results warrant further investigation into the causes of these differences and the appropriateness of using the mixture-averaged diffusion model in the DNS of three-dimensional, premixed turbulent flames at moderate-to-high Karlovitz numbers.

\section*{Acknowledgments}
\label{Acknowledgments}
This material is based upon work supported by the National Science Foundation under Grant Nos.\ 1314109, 1761683, and 1832548.
This research used resources of the National Energy Research Scientific Computing Center, a DOE Office of Science User Facility supported by the Office of Science of the U.S.\ Department of Energy under Contract No.\ DE-AC02-05CH11231, as well as
the Extreme Science and Engineering Discovery Environment (XSEDE), 
which is supported by National Science Foundation grant number 1548562.

\appendix
% Fix for missing space between "Appendix" and letter
\renewcommand*{\thesection}{\appendixname~\Alph{section}}

\section{Availability of material}

The figures in this article, as well as the data and plotting scripts necessary to reproduce them, are available 
openly under the CC-BY license~\cite{repropack}.
Furthermore, the full simulation inputs for and results from NGA are available: 
the three-dimensional multicomponent~\cite{MCdata} and mixture-averaged~\cite{MAdata} hydrogen/air flames,
the two-dimensional hydrogen/air flames and three-dimensional hydrogen/air flame flux quantities~\cite{h2_data},
the \emph{n}-heptane/air flames~\cite{heptane_data},
and the toluene/air flames~\cite{toluene_data}.

%% References can be added with or without bibTeX database
%% Note that the PROCI references style is considered Elsevier non-standard.
%% The original Elsevier bibliography style, elsarticle-num.bst prints paper titles as part of the references, which is different from 
%\bibliography{bibliography.bib}
\bibliography{refs.bib}%%User-specified
\bibliographystyle{elsarticle-num-names.bst}

\end{document}